\documentclass[12pt]{article}
\usepackage{amssymb}
\usepackage{epsfig}

\newcommand{\bq}{\begin{equation}}
\newcommand{\eq}{\end{equation}}
\newcommand{\bqa}{\begin{eqnarray}}
\newcommand{\eqa}{\end{eqnarray}}
\newcommand{\ben}{\begin{enumerate}}
\newcommand{\een}{\end{enumerate}}
\newcommand{\bc}{\begin{center}}
\newcommand{\ec}{\end{center}}

\def\lsim{\lesssim}

\def\pr#1#2#3{ Phys. Rev. ${\bf{#1}}$ (#2) #3}

\def\pl#1#2#3{ Phys. Lett. ${\bf{#1}}$ (#2) #3}
\def\prep#1#2#3{ Phys. Rep. ${\bf{#1}}$ (#2) #3}
\def\np#1#2#3{ Nucl. Phys. ${\bf{#1}}$ (#2) #3}
\def\zp#1#2#3{ Z. f. Phys. ${\bf{#1}}$ (#2) #3}

\def\etal{{\it et.al.\/}}

\global\nulldelimiterspace = 0pt

\def\L{ {\cal L }}

\def\O{ {\cal O }}

\def\t{\hat t}
\def\s{\hat s}
\def\u{\hat u}

\begin{document}
\thispagestyle{empty}
\begin {flushleft}

 PM/99-08\\
%            hep-ph/99....... \\
June 1999\\
\end{flushleft}

\vspace*{2cm}

%---------------------titre ---------------------------------------
\hspace*{-0.5cm}
\begin{center}
{\Large {\bf Precision Tests in $e^+e^-$ and $\gamma\gamma$ processes
and }}\\
{\Large {\bf Virtual New Physics Effects}}
\footnote{{\bf Contribution to 
LP99, International Symposium on Lepton-Photon
Interactions, Stanford, Aug.  1999.}} \\

\hspace*{-0.5cm}
\vspace{1.cm} \\{\bf \large F.M. Renard}\hspace{2.2cm}\null \\
 \vspace{0.2cm}  Physique
Math\'{e}matique et Th\'{e}orique, UMR-CNRS 5825\hspace{2.2cm}\null\\
Universit\'{e} Montpellier
II,  F-34095 Montpellier Cedex 5.\hspace{2.2cm}\null\\[1cm]

\vspace*{0.1cm}
\end{center}
{\bf Summary of works done in collaboration with\\ M. Beccaria,
P. Ciafaloni, D. Comelli, G.J. Gounaris, J. Layssac,\\ P. Porfyriadis,
S. Spagnolo, C. Verzegnassi.}\\

\vspace*{0.2cm}
\begin{center}
{\bf Abstract}\hspace{2.2cm}\null
\end{center}
%
%
%\hspace*{-1.2cm}
\begin{minipage}[b]{16cm}
We report on several works dealing with precison tests of the Standard
Model (SM) and beyond, in the processes $e^+e^-\to f\bar f$ 
($f=$ lepton or quarks) and $\gamma\gamma \to \gamma\gamma,~\gamma Z$. 
We first point out a set of remarkable properties 
of the SM amplitudes and observables at high energies, 
at tree level and at 1-loop, like 
"Sudakov" behaviour or the dominance of purely imaginary 
non-flip amplitudes due to the WWWW box. We then 
consider various types of virtual NP contributions due to supersymmetry,
anomalous gauge boson couplings, technicolour and extended gauge
structures. Effects at LEP2 and LC (in particular with a
laser backscattering 
$\gamma\gamma$ mode) are discussed.
We point out specific features like clear threshold
enhancements.

\end{minipage}

\setcounter{footnote}{0} 
\clearpage
\newpage 
  
\hoffset=-1.46truecm
\voffset=-2.8truecm
\textwidth 16cm
\textheight 22cm
\setlength{\topmargin}{1.5cm}

%\section{} 

In this contribution we report on several works \cite{sud, reval, tool,
light, gZ} dealing with the search for physics beyond the Standard 
Model (SM) that could appear through virtual effects in processes 
measurable at present and future $e^+e^-$
colliders. We successively consider (A) 4-fermion processes  $e^+e^-\to
f\bar f$ ($f=$ lepton or quarks) and (B) photon-photon collisions 
$\gamma\gamma \to \gamma\gamma,~\gamma Z$. In both cases we look for
virtual new physics (NP) effects arising at 1-loop. This should be
especially interesting when no
direct production of the NP degrees of freedom is allowed or detectable
or well identifiable. Each class of processes have specific features
which allow to reach these 1-loop NP effects. In (A) it is the very
high accuracy of the experiments; first at $Z$ peak, where it allows to
establish the SM properties at a very high accuracy; then, at a high
luminosity and a high energy linear collider (LC),
where departures from SM should be enhanced. In (B) it is the fact that
SM contributions only starts at 1-loop, at the same order as the NP
contributions.\\

{\bf (A) Virtual NP effects in $e^-e^-\to f\bar f$ processes}\\

As we just mentioned, the basic ingredient is the measurement of SM
parameters at a very high accuracy at Z peak and at low energy. In
order to exploit this fact in an automatic way and to look for
departures due to NP at higher energies, it is convenient to use the
so-called "Z-peak subtracted representation" \cite{Zsub}. It consists
in using as inputs $\alpha(0)$), $M_Z$ and the 
\underline{$Z$ partial widths and
asymmetries} measured at $Z$ peak. At higher energies the amplitudes 
and the usual observables  are then expressed in terms of sets
of four functions called
$\widetilde{\Delta}_{\alpha,ef}(q^2,\theta)$, $R_{ef}(q^2,\theta)$,
$V^{\gamma Z}_{ef}(q^2,\theta)$, $V^{Z\gamma}_{ef}(q^2,\theta)$.
These functions are
directly associated to subtracted forms ($F_i(q^2)-F_i(M^2_Z)$) of
the $\gamma\gamma$, $ZZ$, $\gamma Z$, $Z
\gamma$ Lorentz structures of the $e^+e^-$ 
annihilation process on which the
1-loop SM or NP contributions are projected \cite{Zsub}. Owing to their
subtracted form, the SM or NP contributions to these functions are
automatically finite and their effects in the various observables
(cross sections $\sigma_f$, forward-backward asymmetries
$A_{FB,f}=\sigma^{FB}_f/\sigma_f$ and polarized asymmetries
$A_{LR,f}={\sigma_{LR,f}/ \sigma_f}~
A^{pol}_{FB,f}={\sigma^{FB}_{LR,f}/ \sigma_f}$) are easily computed
through Born-like expressions. The explicit expressions of the
observables are given in Appendix A. They allow a very fast computation
of any SM or NP effect, analytically or numerically.\\

{\bf --- A first application: the 1-loop pure SM case contributions.}\\

We have collected the full set of 1-loop electroweak SM
contributions to $e^+e^-\to f\bar f$ and their explicit contributions 
to the four functions in \cite{rising}. A
computer code (PALM) has been constructed which also includes the ISR
effects. We have checked that this program which runs in
a very fast way, reproduces at a very good accuracy the
results previously obtained by
existing codes based on conventional representations. The advantage of
our representation is to allow to trace back in a very simple and clear
way each contribution to the various observables. We have then
looked at the high energy behaviour of these 1-loop contributions.
Keeping only their $ln s$ and the $ln^2 s$ terms 
we have established extremely simple expressions for these four
functions and for the corrections to the various observables
\cite{sud}. For example the cross section $\sigma(e^+e^-\to\mu^+\mu^-)$
becomes:

\bqa
\sigma_{\mu}&=&\sigma^{B}_{\mu}[1
+{\alpha(M^2_Z)\over4\pi}\{(7.72N-20.58
)ln{q^2\over M^2_Z}+35.27 ln{q^2\over M^2_W}
-4.59ln^2{q^2\over M^2_W}
\nonumber\\
&&
+4.79ln{q^2\over M^2_Z} -1.43ln^2{q^2\over M^2_Z}\}+........]
\label{sigmu}\eqa
\noindent
where $\sigma^{(B)}_{\mu}=0.11106/q^2 (fb/TeV^2)$.
The first $ln{q^2\over M^2_Z}$ term corresponds to the Renormalization
Group (RG) contribution (self-energy diagrams and universal Bosonic
triangles; $N=3$ is the number of fermionic generations), 
whereas the other (linear and quadratic) logarithmic terms
are of "Sudakov" type. The $ln^2 s$ terms come from well 
identified specific triangle diagrams with one gauge boson exchanged and
from box diagrams with a pair of gauge bosons.
Similar expressions for the other muonic or hadronic observables
can be found in \cite{sud}.\par
As one can see in Fig.1, the RG linear logarithmic contribution is
smaller than the Sudakov one. Although the linear and quadratic Sudakov
contributions have opposite signs, the total effect is much different
from the RG one, so that ignoring them would give a completely wrong
result. One also sees that in the multi-TeV range the individual
contributions reach a relative magnitude of the 10\% level thus raising
the question of possible non negligible 2-loop effects.\par

The dots that appear in the brackets of eqs.(\ref{sigmu}) 
refer to the "non-leading" terms. These could 
either be constants or $O({1\over q^2})$ components 
whose asymptotic effect vanishes. We have checked that indeed, in the
TeV region, these
asymptotic results agree at the permille level
with exact SM predictions obtained 
from the semianalytic program (PALM)\cite{rising} and from TOPAZ0
\cite{Topaz0}.
At non asymptotic 
energies, for example in the range $0.2~TeV <\sqrt{s}< 1~TeV$, one can
globally replace the dots by
fitted constants. In the case of $\sigma_{\mu}$ the dots can be
replaced by $-13.5{\alpha(M^2_Z)\over\pi}$, which optimizes the
approximation at $0.5~TeV$. \par
Eq.(\ref{sigmu}) and similar ones obtained in
ref.\cite{sud} provide very good descriptions of the various
polarized and unpolarized observables in $e^+e^-\to f\bar f$
at high energy. They can be useful when discussing the potential of
future colliders for testing the Standard Model predictions and for
looking at departures due to NP contributions.\\

{\bf --- Supersymmetric effects.}\\

The second application concerns Supersymmetry effects. We have added
to SM diagrams the corresponding sets of self-energy, vertex and box
diagrams with supersymmetric
partners, sfermions and gauginos.
The results now depend on the supersymmetric model and the
corresponding parameters which are used. We have
considered the MSSM \cite{MSSM}
(whose parameters are $M_1.~M_2,~\mu,~tan\beta)$
with the
GUT relation between the SU(2) $\otimes$ U(1) gauginos
soft mass parameters $M_1=\frac{5}{3}\tan^2\theta_w M_2$. 
We have first looked at the possibility of a signal in 
$e^+e^-\to f\bar f$ at LEP2 due to a light chargino \cite{reval}.

For illustration we fix the light chargino mass at
105~GeV, the sleptons physical masses at 120~GeV and the squarks
physical ones at 200~GeV.  We set $\tan\beta=1.6$, and verified
that varying it from 1.6 to 40 does not produce any appreciable change. With
this choice, we computed the \underline{relative} SUSY shifts on the three
chosen observables ${\cal O}_i$, $\Delta^{\rm SUSY}{\cal O}\equiv \frac{{\cal
O}^{\rm SUSY}-{\cal O}^{\rm SM}}{{\cal O}^{\rm SM}}$ (${\cal O}_{1, 2,
3}=\sigma_\mu, \sigma_5, A_{FB, \mu}$).

Fig.~(\ref{scan105}) shows the variations of the relative effects on the
observables when $\sqrt{q^2}=200$~GeV and 
$\mu$ varies in its allowed range. One sees that the size of
the SUSY contribution to the muon asymmetry remains systematically negligible, 
well below the six-seven permille limit that represents an
optimistic experimental reach in this case~\cite{LEP2}.
On the contrary, in
the case of the muon and hadronic cross sections, the size of the effect
approaches, for large $|\mu|$ values, a limit of six permille in $\sigma_\mu$ 
and 
four permille in $\sigma_5$ that
represent a conceivable experimental reach, at the end of the overall LEP2 
running
period.

This explains in fact our choice of the value $M_{\chi^+_{light}}=
105$~GeV with LEP2 limit at 200~GeV; other couples of the light chargino mass
and of the LEP2 limit separated by a larger gap would produce a smaller
effect, i.e. an unobservable one. On the other hand, smaller gaps (e.g. a
lighter but still unproduced chargino \underline{or} a larger LEP2 limit)
would increase the effect, towards the one percent 
values that appear to be experimentally realistic.

We have analyzed various possibilities, $|\mu|>>M_2$ (``gaugino
like''), $M_2>>|\mu|$ (``higgsino like''). We present 
in Fig.~(\ref{lightchar}) the energy behaviour of the three main
unpolarized observables in the high $|\mu|$ case, showing a 
remarkable threshold enhancement.

We have also considered the different situation, 
where the lightest chargino is
``heavy'' and decoupled, setting its mass equal to 300 GeV, and assuming that
all sleptons are now ``light'' (i.e.  $m_{\tilde{l}}=$ 105 GeV).
The signal has almost completely
disappeared. This is the consequence of the 
P-wave depression factor $\approx q^2-4
m_{\tilde{l}}^2$ which appears for spinless sfermions and 
which washes out the threshold enhancement \cite{reval}.\\

Applications to a $500~GeV$ LC are in progress.\\

{\bf --- Residual NP effects.}\\

We have considered NP effects leading to anomalous gauge boson
couplings. This type of effects is usually described by a set
of $SU(2)\times U(1)$ gauge invariant operators. We have chosen the
linear representation and the set of $dim=6$ operators \cite{Hag}.
Among them a subset dubbed "blind" is constrained
by direct measurements in $e^+e^-\to W^+W^-$ and another set
dubbed "superblind" by $e^+e^-\to HZ,~ H\gamma$ \cite{LEP2}. 
The remaining
"non-blind" set of 4 operators $\O_{DW},~ \O_{DB},~  \O_{BW},~  
\O_{\Phi 1}$ 
is responsible for modifications 
of the gauge boson propagators:

\bq
\L^{(NB)}= {f_{DW}\over\Lambda^2}\O_{DW}+{f_{DB}\over\Lambda^2}\O_{DB}
+{f_{BW}\over\Lambda^2}\O_{BW}+{f_{\Phi,1}\over\Lambda^2}\O_{\Phi,1}
\label{L}\eq
\bqa
\O_{DW} & =& Tr ([D_{\mu},{\overrightarrow{W}}_{\nu
\rho})] [D^{\mu},{\overrightarrow{W}}^{\nu \rho}])  \ \ \
  , \ \  \label{listDW}  \\[0.1cm]
\O_{DB} & = & -{g'^2\over2}(\partial_{\mu}B_{\nu \rho})
(\partial^\mu B^{\nu
\rho}) \ \ \  , \ \   \label{listDB} \\[0.1cm] 
\O_{BW} & =&  \Phi^\dagger {B}_{\mu \nu}
\overrightarrow \tau \cdot {\overrightarrow{W}}^{\mu \nu} \Phi
\ \ \  , \ \  \label{listBW}  \\[0.1cm] 
\O_{\Phi 1} & =& (D_\mu \Phi^\dagger \Phi)( \Phi^\dagger
D^\mu \Phi) \ \ \  , \ \       \label{listPhi1}   
\eqa

They have already been constrained
by measurements of $e^+e^-\to f \bar f$ at $Z$ peak \cite{Hag}, 
but it has
been shown that improvements can be expected by measurements
of $e^+e^-\to f \bar f$  cross sections 
at higher energies\cite{clean,
HagLC}. Using the virtues of the
$Z$ peak subtracted representation, we have shown that the largest
improvements expected from LEP2 or LC measurements will
occur for the operators $\O_{DW},~\O_{DB}$ which
lead to strong energy dependent effects. In a new analysis
\cite{bound} including
low energy and $Z$ peak data together with expectations at LEP2 
we have obtained the results shown in Fig.\ref{low-high}, 
which for example
amount to improve the limits as follows:\\

\begin{center}

\begin{tabular}{c|cccc}
     & $\delta f_{DW}$  & $\delta f_{BW}$       & $\delta f_{DB}$       &
$\delta f_{\Phi, 1}$ \\ 
\hline \\
{\bf Low}  & 0.28 & 1.43 & 6.27 & 0.088 \\
{\bf Low+High} & 0.18 & 0.32 & 1.15 & 0.035 
\end{tabular}
\end{center}

{\bf Table1}: Bounds on the anomalous gauge
couplings obtained with a combined fit of present and 
future experimental
data. The defintion of the parameter uncertainties 
adopted here is the 1
$\sigma$ error in the $\chi^2$ minimization. 
{\bf Low} refers to the results from LEP1, APV and 
from the measurement of 
$M_W$, {\bf High} to the cross-section and asymmetry
 measurements at LEP2. 

\vspace{0.5cm}

At a 500 GeV linear collider, with a luminosity of
50 $fb^{-1}$ the sensitivity is expected to improve
at the level of $\delta f_{DW}=0.06$, $\delta f_{BW}=0.27$,
$\delta f_{DB}=0.22$, $\delta f_{\Phi, 1}=0.04$ in \cite{HagLC}
and $\delta f_{DW}=0.025$, $\delta f_{DB}=0.16$ 
with unpolarized beams, and 
$\delta f_{DW}=0.014$, $\delta f_{DB}=0.08$ with polarized beams
in \cite{RVpol}.  With a higher LC luminosity \cite{LC}, these
numbers should scale like $\L_{ee}^{-{1\over2}}$.\\

{\bf --- Other types of NP.}\\ 

We close this subsection by mentioning that 
the "$Z$-peak subracted representation" has also been used for
establishing constraints on the effects of Higher Vector Bosons,
either of gauge nature (for example $Z'$) or of composite nature
(for example Technicolour resonances). See ref.\cite{Zsub, TCchiv}
 where
limits for masses and couplings were given.\\

 {\bf (B) Virtual NP effects in $\gamma\gamma \to \gamma\gamma,~\gamma
Z,~ZZ$ processes}\\

These processes are particularly interesting because the SM contribution
only starts at 1-loop\cite{Jikia};
a contrario $\gamma\gamma \to W^+W^-$ has
tree level contributions. Therefore
these neutral processes provide a clean window
to new physics. The laserbackscattering device should allow to provide
intense and high energy photon beams at an $e^+e^-$ linear collider.
For example the TESLA project \cite{LC} announce the possibility 
of accumulating an $e^+e^-$ luminosity of
about $1000~fb^{-1}$ in two or three years. The
photon fluxes $dL_{\gamma\gamma}/d\tau$ that multiply the 
$e^+e^-$ luminosity are expected to be of the order
of 1 or even more for the range
$E^{\gamma\gamma}_{cm} \lsim 0.8~ E^{ee}_{cm}$ \cite{LCgg}. 
This should allow to
make precise measurements of these processes whose cross sections are of
the order of a few tens of $fb$. NP effects leading to 
departures from SM at the level of 
1 percent should then be observable.\\
A special feature of the SM amplitudes in the above processes is
the dominance of a few purely imaginary non flip
amplitudes \cite{tool}; for example in $\gamma\gamma\to\gamma\gamma$:

\bqa
 F_{\pm\pm\pm\pm}(\hat{s},\hat{t},\hat{u})
& \simeq &  -i~ 16\pi\alpha^2\Bigg [{\hat{s}\over \hat{u}}
Ln\Big |{\hat{u}\over M^2_W}\Big | +
{\hat{s}\over \t} Ln\Big |{\hat{t}\over M^2_W}\Big |\Bigg ]
\ \ , \label{pppp} \\
 F_{\pm\mp\pm\mp}(\hat{s},\hat{t},\hat{u})
& \simeq &
-i~ 12\pi\alpha^2{\s-\t\over \u}+i{8\pi\alpha^2\over
\hat{u}^2}(4\hat{u}^2-3\hat{s}\hat{t})
\Bigg [Ln\Big |{\hat{t}\over \hat{s}}\Big |\Bigg ]\nonumber\\
&&  -i~ 16\pi\alpha^2 \Bigg [{\hat{u}\over \hat{s}}
Ln \Big |{\hat{u}\over M^2_W}\Big | + {\hat{u}^2\over \hat{s}\hat{t}}
Ln \Big |{\hat{t}\over M^2_W} \Big |\Bigg ]
\label{pmpm} \ ,
\eqa

Similar amplitudes appear in $\gamma\gamma \to \gamma Z$ with a
coupling factor $c_W/s_W$ and $M^2_Z/s$ corrections \cite{gZ}.
In the high energy limit, the dominant 1-loop SM contributions satisfy
the relations

\bq
F^W_{\gamma\gamma\to\gamma Z}
\simeq {c_W\over s_W} F^W_{\gamma\gamma\to\gamma \gamma}
\eq
\noindent

\bq
F^f_{\gamma\gamma\to\gamma Z}
\simeq {g^Z_{Vf}\over Q_f} F^f_{\gamma\gamma\to\gamma \gamma}
\eq
\noindent
with, for a standard fermion $f$,

\bq
g^Z_{Vf} = {t^f_3 - 2Q_f s^2_W\over 2c_Ws_W}
\eq

 The dominance of imaginary parts
is due to the large W box contribution. This is confirmed by an
exact numerical computation of all SM amplitudes as we can see in
Fig.\ref{ampSM}. This situation
is reminiscent of the Pomeron dominance in VDM processes at the
hadronic GeV scale.\par
Non standard effects due to the interference with SM 
will therefore appear predominantly when the
amplitudes have large imaginary parts. Typical examples
are threshold enhancements, resonances, unitarity saturating
phenomena \cite{tool, TC1, sat}.
 A few illustrations are reproduced
in Fig.\ref{models}.
On the opposite, note that residual NP effects described by effective
lagrangians giving real amplitudes, are not favored.\par
Beam polarization should allow to make checks of the nature of the NP
effects. Assuming 80 \% electron beam polarization and fully polarized
laser beams,
polarized photon-photon fluxes are expected to be large \cite{Tsi,
light}. They allow to consider 8 different types of quantities
in $\gamma\gamma \to \gamma Z$ (they reduce to 6 in
$\gamma\gamma \to \gamma \gamma$)
denoted $\bar{\sigma}_{ij}$ and listed in Appendix B (see
also \cite{light}
and in \cite{gZ}).\par 
We have developed the study of supersymmetric effects, considering
contributions from fermionic and scalar partners. In Fig
\ref{SUSYgg}-\ref{SUSYgZ1}, we have illustrated the
$\gamma\gamma\to\gamma\gamma,~Z$ cases with one gaugino and one slepton.
As the corresponding  (gaugino, slepton)
box amplitudes in
$\gamma\gamma\to\gamma\gamma$ depend on the coupling factor
$Q^4$, the result equally apply to any other (fermion, scalar) with the
same mass. So this process provides 
model independent tests. In addition it is experimentally clean. 
As opposed to real production of new particles
this search for virtual effects is not affected
by the complexity of the decay modes.\\
The $\gamma\gamma\to\gamma Z$ process is also clean; 
all Z modes (trigerred by the associated photon) can be used.
The rate is about 6 times
larger than the $\gamma\gamma\to\gamma\gamma$ one. 
The couplings factor in the box due a particle X 
is now $Q^3_X g^Z_{VX}$. Its presence 
should allow, first to confirm a possible effect in 
$\gamma\gamma\to\gamma\gamma$, secondly to
disentangle different possibilities \cite{abdel} which differ by
the magnitude like chargino/higgsino, or even the sign
of $g^Z_{VX}$ like slepton-L/slepton-R,
and lead, through the interference term, to corresponding
departures from SM predictions.\par
Similar applications can be done to other types of virtual
contributions, for example new fermions and bosons in Technicolour
schemes \cite{TC1}.\\
The process $\gamma\gamma\to ZZ$ should give further informations
(in particular on the Higgs sector) 
and its study is in progress.\\

{\Large \bf Conclusions}

We have pointed out several remarkable properties of the processes
$e^+e^-\to f\bar f$ 
and $\gamma\gamma \to \gamma\gamma,~\gamma Z$ at high energies and
their implications for the search of virtual NP effects. This search
requires the availability of high energy and high luminosity
$e^+e^-$ and  $\gamma\gamma$ colliders.

There are historical examples about the way precison tests 
can give hints about new particles and interactions.
Recently, at $Z$ peak, we had several examples with neutrino counting,
the excellent hint for the top quark and maybe now a hint for a
light Higgs boson.\par
At higher energies the search for NP in standard processes
may proceed in a similar
way, in the sense that hints may come from precision
measurements revealing the presence of some anomalous
contribution. The high accuracy at which the SM parameters
have been measured at Z peak and at low energies, put together with
accurate measurements at high energies should be the clue to this
type of searches.\par
 The "$Z$-peak subtracted representation", that
we have emphasized in this report, is especially suitable for the
studies in $e^+e^-\to f\bar f$ at high energies,
as it takes into account in an automatic
way, the measurements at $Z$ peak and as it singles out the
departures due to new contributions rising with the energy.
We have treated several examples taken from NP predicted by
Supersymmetry or Technicolour models.\par
In $\gamma\gamma \to \gamma\gamma,~\gamma Z$ processes the 
situation is somewhat different because the SM contribution
and the NP contribution both start at 1-loop. This privileged
situation of NP effects may compensate the weakness of the
corresponding cross sections.\par
For example we have shown, both in $e^+e^-$ and in $\gamma\gamma $
collisions that NP signals may come in standard processes
from visible "threshold
enhancements".
This method is independent of the one which consists in
looking at the direct production of new particles and studying
their  decay modes.
It  should be especially advantageous for the search of
new particles decaying through a long chain of processes, which are
difficult to extract from a background (ex. gauginos, sleptons or
PGB's).\par
Another remark is that in $\gamma\gamma \to \gamma\gamma$ the
unpolarized cross section (because of the $Q^4$ box contribution)
cumulates positive contributions from NP above threshold. Thus, 
in the high energy limit,
this cross section provides a kind of counting of the number of states
involved in the loop.
For example, if  SUSY is realized in Nature below the
TeV-scale, then it would be quite plausible  that a
chargino, as well as all six charged sleptons  and the $\tilde
t_1$ squark, lie in a limited  mass range.
  In such a case, a clear signal could be observable.\par
The process $\gamma \gamma \to \gamma Z$ with a larger cross section
and the presence of the $Z$ coupling as well
as the
availability of polarized $\gamma\gamma$ collisions
should give additional informations about the nature of 
the particles produced.\par
We therefore conclude, that important physical information
could arise from the study of  the $e^+e^- \to f\bar f$ and the
$\gamma \gamma \to \gamma \gamma,~\gamma Z$ processes, and that this
certainly constitutes
an  argument favoring the availability of a high energy, high
luminosity Linear Collider and its 
laser $\gamma \gamma $ option.

\newpage

{\large \bf Appendix A: Expressions of $e^+e^-\to f\bar f$ observables
in the  $Z$-peak subtracted representation.}\\

The unpolarized cross section for $e^+e^-\to f\bar f$

\bq 
\sigma_f =  {4\pi N_f q^2\over3} \int^{+1}_{-1} dcos\theta ~~ 
[{3\over8}(1+cos^2\theta)U_{11}+{3\over4}cos\theta~ U_{12}] 
\eq
\noindent
and the forward-backward asymmetry

\bq 
\sigma^{FB}_f =  {4\pi N_f q^2\over3}\{ \int^{+1}_{0} - \int^{0}_{-1}\}
dcos\theta ~~ 
[{3\over8}(1+cos^2\theta)U_{11}+{3\over4}cos\theta~ U_{12}] 
\eq

\bq
A_{FB,f}=\sigma^{FB}_f/\sigma_f
\eq
\noindent
are expressed in terms of $U_{11}$ and $U_{12}$:

\bqa  
U_{11}=&&
{\alpha^2(0)Q^2_f\over q^4}[1+2\delta\tilde{\Delta}^{(lf)}\alpha(q^2)]
\nonumber\\
&&+2[{\alpha(0)|Q_f|}]{q^2-M^2_Z\over
q^2((q^2-M^2_Z)^2+M^2_Z\Gamma^2_Z)}[{3\Gamma_l\over
M_Z}]^{1/2}[{3\Gamma_f\over N_b M_Z}]^{1/2}
{\tilde{v}_l \tilde{v}_f\over
(1+\tilde{v}^2_l)^{1/2}(1+\tilde{v}^2_f)^{1/2}}\nonumber\\
&&\times[1+
\tilde{\Delta}^{(lf)}\alpha(q^2) -R^{(lf)}(q^2)
-4s_lc_l
\{{1\over \tilde{v}_l}V^{(lf)}_{\gamma Z}(q^2)+{|Q_f|\over\tilde{v}_f}
V^{(lf)}_{Z\gamma}(q^2)\}]\nonumber\\ 
&&+{[{3\Gamma_l\over
M_Z}][{3\Gamma_f\over N_f M_Z}]\over(q^2-M^2_Z)^2+M^2_Z\Gamma^2_Z}
\nonumber\\
&&\times[1-2R^{(lf)}(q^2)
-8s_lc_l\{{\tilde{v}_l\over1+\tilde{v}^2_l}V^{(lf)}_{\gamma
Z}(q^2)+{\tilde{v}_f|Q_f|\over(1+\tilde{v}^2_f)}
V^{(lf)}_{Z\gamma}(q^2)\}]\ , 
\label{U11pro}
 \\[0.2cm]
U_{12}=&& 2[{\alpha(0)|Q_f|}]{q^2-M^2_Z\over
q^2((q^2-M^2_Z)^2+M^2_Z\Gamma^2_Z)}
[{3\Gamma_l\over M_Z}]^{1/2}[{3\Gamma_f\over N_f
M_Z}]^{1/2}{1\over(1+\tilde{v}^2_l)^{1/2}(1+\tilde{v}^2_f)^{1/2}}
\nonumber\\
&&\times[1+
\tilde{\Delta}^{(lf)}\alpha(q^2) -R^{(lf)}(q^2)]\nonumber\\
&&+{[{3\Gamma_l\over M_Z}][{3\Gamma_f\over N_f
M_Z}]\over(q^2-M^2_Z)^2+M^2_Z\Gamma^2_Z}
[{4\tilde{v}_l \tilde{v}_f\over(1+\tilde{v}^2_l)(1+\tilde{v}^2_f)}]
\nonumber\\
&&\times[1-2R^{(lf)}(q^2)-4s_lc_l
\{{1\over \tilde{v}_l}V^{(lf)}_{\gamma Z}(q^2)+{|Q_f|\over\tilde{v}_f}
V^{(lf)}_{Z\gamma}(q^2)\}] \ , 
\label{U12pro}
 \\[0.2cm]
\eqa

\noindent
In the left-right polarized case, we have the two other combinations
$U_{21}$ and $U_{22}$:\\

\bq
\sigma_{LR,f} = {4\pi N_f q^2\over3} \int dcos\theta ~~
[{3\over8}(1+cos^2\theta)U_{21}+{3\over4}cos\theta~ U_{22}]
\eq

\bq 
\sigma^{FB}_{LR,f} =  {4\pi N_f q^2\over3}
\{ \int^{+1}_{0} - \int^{0}_{-1}\}
dcos\theta ~~ 
[{3\over8}(1+cos^2\theta)U_{21}+{3\over4}cos\theta~ U_{22}] 
\eq

\bq
A_{LR,f}={\sigma_{LR,f}/ \sigma_f}~~~~~~~~~~
A^{pol}_{FB,f}={\sigma^{FB}_{LR,f}/ \sigma_f}
\eq

\bqa
U_{21}=&& 2[{\alpha(0)|Q_f|}]{q^2-M^2_Z\over
q^2((q^2-M^2_Z)^2+M^2_Z\Gamma^2_Z)}
[{3\Gamma_l\over
M_Z}]^{1/2}[{3\Gamma_f\over N_f
M_Z}]^{1/2}{\tilde{v}_f\over(1+\tilde{v}^2_l)^{1/2}
(1+\tilde{v}^2_f)^{1/2}}\nonumber\\
&&\times[1+\tilde{\Delta}^{(lf)}\alpha(q^2) -R^{(lf)}(q^2)
-{4s_lc_l|Q_f|\over\tilde{v}_f}V^{(lf)}_{Z\gamma}(q^2)]
+{[{3\Gamma_l\over
M_Z}][{3\Gamma_f\over N_f
M_Z}]\over(q^2-M^2_Z)^2+M^2_Z\Gamma^2_Z}
[{2\tilde{v}_l \over(1+\tilde{v}^2_l)}]\nonumber\\
&&\times[1-2R^{(lf)}(q^2)-4s_lc_l
\{{1\over \tilde{v}_l}V^{(lf)}_{\gamma
Z}(q^2)+{2\tilde{v}_f|Q_f|\over(1+\tilde{v}^2_f)}
V^{(lf)}_{Z\gamma}(q^2)\}]  \ , 
\label{U21pro}
 \\[0.2cm]
U_{22}= && 2[{\alpha(0)|Q_f|}]{q^2-M^2_Z\over
q^2((q^2-M^2_Z)^2+M^2_Z\Gamma^2_Z)}
[{3\Gamma_l\over
M_Z}]^{1/2}[{3\Gamma_f\over N_f
M_Z}]^{1/2}{\tilde{v}_l\over(1+\tilde{v}^2_l)^{1/2}
(1+\tilde{v}^2_f)^{1/2}}\nonumber\\
&&\times[1+\tilde{\Delta}^{(lf)}\alpha(q^2) -R^{(lf)}(q^2)
-{4s_lc_l\over \tilde{v}_l}V^{(lf)}_{\gamma Z}(q^2)]
+{[{3\Gamma_l\over
M_Z}][{3\Gamma_f\over N_f
M_Z}]\over(q^2-M^2_Z)^2+M^2_Z\Gamma^2_Z}
[{2\tilde{v}_f \over(1+\tilde{v}^2_f)}]\nonumber\\
&&\times[1-2R^{(lf)}(q^2)-4s_lc_l
\{{2\tilde{v}_l\over(1+\tilde{v}^2_l)}
V^{(lf)}_{\gamma Z}(q^2)+{|Q_f|\over
\tilde{v}_f}V^{(lf)}_{Z\gamma}(q^2)\}] \ . \label{U22pro}
\eqa

$N_f$ is the conventional color factor which contains 
standard QCD corrections, the inputs are $\alpha(0)$, $M_Z$ and
$\Gamma_l$, $\Gamma_f$, $\tilde{v}_e \equiv 1-4s^2_l$, 
$\tilde{v}_f\equiv 1-4|Q_f|s^2_f$ measured at $Z$ peak.\par
This applies to $e^+e^- \to leptons$ (f=l) and to $e^+e^- \to hadrons$ 
(summing $\sum_f\sigma_f$
and $\sum_f\sigma_{LR,f}$ over $f=2u+2d+b$).\par
Each application then consists in considering contributions
to the four functions $\tilde{\Delta}^{(lf)}\alpha(q^2)$,
$R^{(lf)}(q^2)$, $V^{(lf)}_{\gamma Z}(q^2)$ and 
$V^{(lf)}_{Z\gamma}(q^2)$. These are obtained by projecting the
considered SM or NP amplitude on the $\gamma\gamma$, $ZZ$, $\gamma Z$
and $Z\gamma$ Born Lorentz structures \cite{Zsub}.\par 
For example in the case of universal modifications to the gauge boson
propagators due to the $dim=6$ operators $\O_{DW},~ \O_{DB}$ one gets
\cite{bound}:

\bqa
\tilde{\Delta}^{(AGC)}_{\alpha} (q^{2}) =
-8 \pi \alpha \, \frac{q^{2}}{\Lambda^{2}} \,
\left[ f_{DW} + f_{DB} \right]\label{da}
\eqa

\bqa
R^{(AGC)}(q^{2}) = 8 \pi \alpha \, \frac{(q^{2}-M^{2}_{z})}{\Lambda^{2}}
\,
\left[ \frac{c^2_l}{s^2_l} \, f_{DW} +
\frac{s^2_l}{c^2_l} \, f_{DB} \right ]
\label{R}
\eqa

\bqa
V^{(AGC)}(q^{2}) = 8 \pi \alpha \, \frac{(q^{2}-M^{2}_{z})}{\Lambda^{2}}
\,
\left[ \frac{c_l}{s_l} \, f_{DW} -
\frac{s_l}{c_l} \, f_{DB} \right ]
\label{V}
\eqa

\newpage

{\large \bf Appendix B: General form of $\gamma\gamma\to \gamma\gamma
~,\gamma Z$ cross sections.}\\

The possibility to use polarized or unpolarized
$\gamma\gamma$ collisions in an LC operated in the $\gamma
\gamma $ mode,  through laser backscattering \cite{LC, LCgg}
is described
in  \cite{Tsi, light}.  The assumption of
Parity invariance leads to the following form for the
$\gamma \gamma \to \gamma \gamma,~\gamma Z $
cross section (note that a factor ${1\over2}$ should be applied in the
$\gamma \gamma \to \gamma \gamma$ case)
\bqa
{d\sigma\over d\tau d\cos\vartheta^*}&=&{d \bar L_{\gamma\gamma}\over
d\tau} \Bigg \{
{d\bar{\sigma}_0\over d\cos\vartheta^*}
+\langle \xi_2 \xi_2^\prime 
\rangle{d\bar{\sigma}_{22}\over d\cos\vartheta^*}
+[\langle\xi_3\rangle\cos2\phi{d\bar{\sigma}_{3}\over d\cos\vartheta^*}
+\langle\xi_3^ \prime\rangle\cos2\phi^\prime
{d\bar{\sigma'}_{3}\over d\cos\vartheta^*}]
\nonumber\\
&&+\langle\xi_3 \xi_3^\prime\rangle[{d\bar{\sigma}_{33}\over d\cos\vartheta^*}
\cos2(\phi+\phi^\prime)
+{d\bar{\sigma}^\prime_{33}\over
d\cos\vartheta^*}\cos2(\phi- \phi^\prime)]\nonumber\\
&&+[\langle\xi_2 \xi_3^\prime\rangle\sin2 \phi^\prime
{d\bar{\sigma}_{23}\over d\cos\vartheta^*}-
\langle\xi_3 \xi_2^\prime\rangle\sin2\phi
{d\bar{\sigma'}_{23}\over d\cos\vartheta^*}] \Bigg \} \ \ ,
\label{sigpol}
\eqa

where
\bqa
{d\bar \sigma_0\over d\cos\vartheta^*}&=&
\left ({\beta_Z\over64\pi\hat{s}}\right )
\sum_{\lambda_3\lambda_4} [|F_{++\lambda_3\lambda_4}|^2
+|F_{+-\lambda_3\lambda_4}|^2] ~ ,  \label{sig0} \\
{d\bar{\sigma}_{22}\over d\cos\vartheta^*} &=&
\left ({\beta_Z\over64\pi\hat{s}}\right )\sum_{\lambda_3\lambda_4}
[|F_{++\lambda_3\lambda_4}|^2
-|F_{+-\lambda_3\lambda_4}|^2]  \ , \label{sig22} \\
{d\bar{\sigma}_{3}\over d\cos\vartheta^*} &=&
\left ({-\beta_Z\over32\pi\hat{s}}\right ) \sum_{\lambda_3\lambda_4}
Re[F_{++\lambda_3\lambda_4}F^*_{-+\lambda_3\lambda_4}]  \ ,
\label{sig3} \\
{d\bar{\sigma'}_{3}\over d\cos\vartheta^*} &=&
\left ({-\beta_Z\over32\pi\hat{s}}\right ) \sum_{\lambda_3\lambda_4}
Re[F_{++\lambda_3\lambda_4}F^*_{+-\lambda_3\lambda_4}]  \ ,
\label{sig3p} \\
{d\bar \sigma_{33} \over d\cos\vartheta^*}& = &
\left ({\beta_Z\over64\pi\hat{s}}\right ) \sum_{\lambda_3\lambda_4}
Re[F_{+-\lambda_3\lambda_4}F^*_{-+\lambda_3\lambda_4}] \ ,
\label{sig33} \\
{d\bar{\sigma}^\prime_{33}\over d\cos\vartheta^*} &=&
\left ({\beta_Z\over64\pi\hat{s}}\right ) \sum_{\lambda_3\lambda_4}
Re[F_{++\lambda_3\lambda_4}F^*_{--\lambda_3\lambda_4}] \  ,
\label{sig33prime} \\
{d\bar{\sigma}_{23}\over d\cos\vartheta^*}& = &
\left ({\beta_Z\over64\pi\hat{s}}\right ) \sum_{\lambda_3\lambda_4}
Im[F_{++\lambda_3\lambda_4}F^*_{+-\lambda_3\lambda_4}] \ ,
\label{sig23}\\
{d\bar{\sigma'}_{23}\over d\cos\vartheta^*}& = &
\left ({\beta_Z\over64\pi\hat{s}}\right ) \sum_{\lambda_3\lambda_4}
Im[F_{++\lambda_3\lambda_4}F^*_{-+\lambda_3\lambda_4}] \ ,
\label{sig23p}
\eqa
are expressed in terms of the $\gamma \gamma \to \gamma \gamma,
~\gamma Z$
amplitudes given in
Appendix A of \cite{light}, \cite{gZ}.\\
 $\beta_Z=1-{M^2_Z\over \hat{s}}$ and
$\vartheta^*$ is the scattering angle in
the $\gamma \gamma $ rest frame;
$\tau \equiv s_{\gamma \gamma}/s_{ee}$.
Note that only $d\bar \sigma_0/ d\cos\vartheta^*$
is positive definite and that 
${d\bar{\sigma'}_{3}\over d\cos\vartheta^*}(\cos\vartheta^*)=
{d\bar{\sigma}_{3}\over d\cos\vartheta^*}(-\cos\vartheta^*)$, and
${d\bar{\sigma'}_{23}\over d\cos\vartheta^*}(\cos\vartheta^*)$ $ =
{d\bar{\sigma}_{23}\over d\cos\vartheta^*}(-\cos\vartheta^*)$, which
are identical in the case of the $\gamma\gamma$ final state.\par

A quick estimate of the unpolarized cross section
can be done using the helicity amplitudes 
$F_{\lambda_1\lambda_2\lambda_3\lambda_4}$ mentioned in the text,
eq.(5-7).\\
For a complete description one needs to take into account subleading
amplitudes \cite{light, gZ}.\\
${d \bar L_{\gamma\gamma}\over d\tau}$ is the $\gamma\gamma$ flux, 
and $\langle\xi_i \rangle$
$\langle\xi_i \xi_j^\prime\rangle$ are average stokes vectors
describing the polarization state of the backscattered photon
(possibly linearly polarized along the direction defined by the angle
$\phi$ or $\phi^\prime$, see \cite{light, gZ}). These
are computable simple functions of 
$\tau\equiv s_{\gamma\gamma}/s_{ee}$ and of the polarization degrees
of the $e^{\pm}$ and laser beams, given in  ref.\cite{LCgg, light}.

\begin{figure}[p]
\[
\hspace{-2.5cm}
\epsfig{file=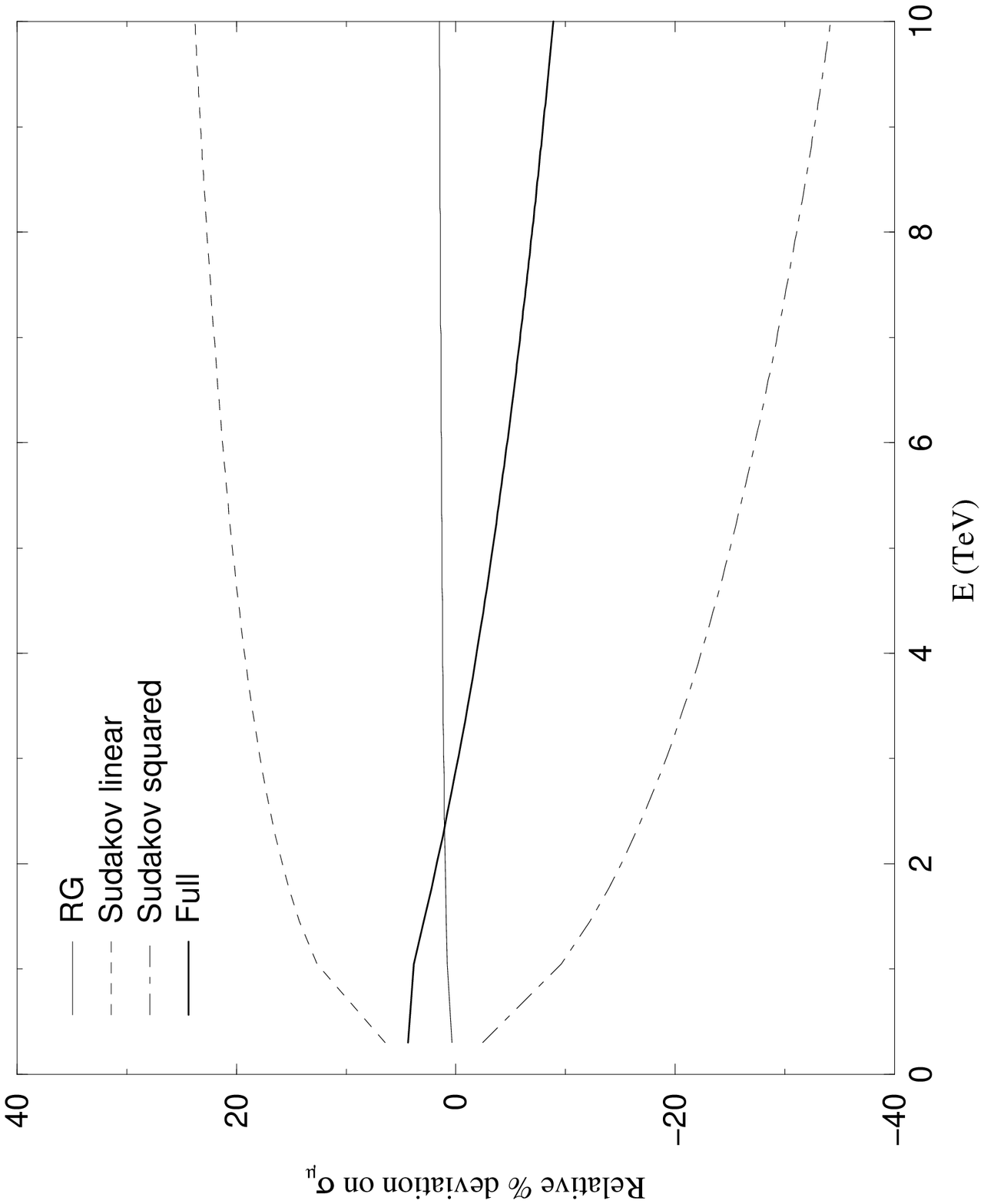,height=18cm}
\]
%\vspace*{-19cm}
\caption[5]{ Logarithmic 1-loop Standard Model
contributions to the asymptotic cross
section $\sigma(e^+e^-\to\mu^+\mu^-)$.}
\label{Fig5}
\end{figure}

\begin{figure}[htb]\setlength{\unitlength}{1cm}
\center{\epsfig{ file=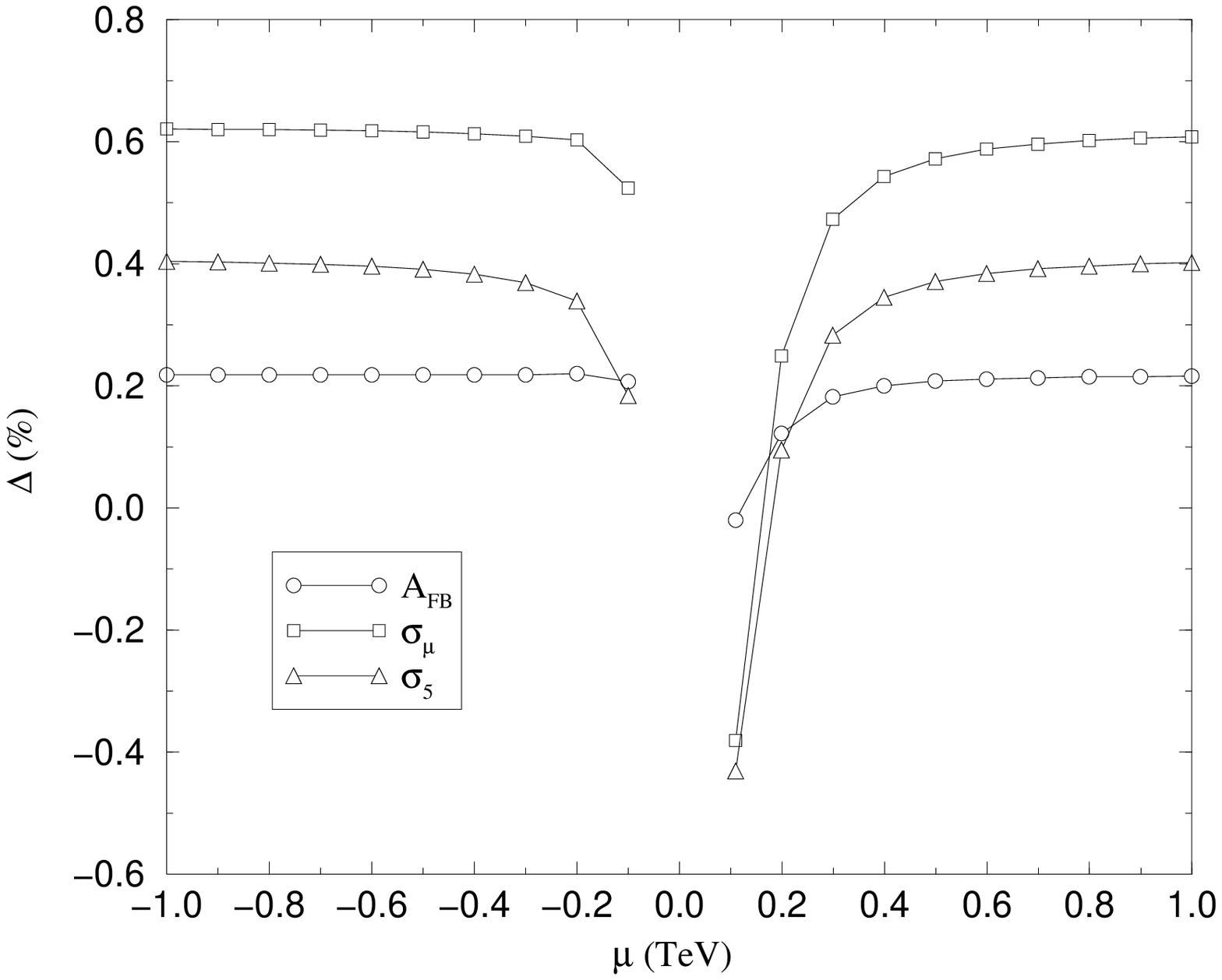,
width=10cm}}
\caption{SUSY effects on the three considered observables with the mass of the
lightest chargino fixed at 105 GeV and $\tan\beta=1.6$.  $m_{\tilde{q}}$ is
fixed at 200 GeV and $m_{\tilde{l}}$ at 120 GeV.}
\label{scan105}
\center{\epsfig{ file=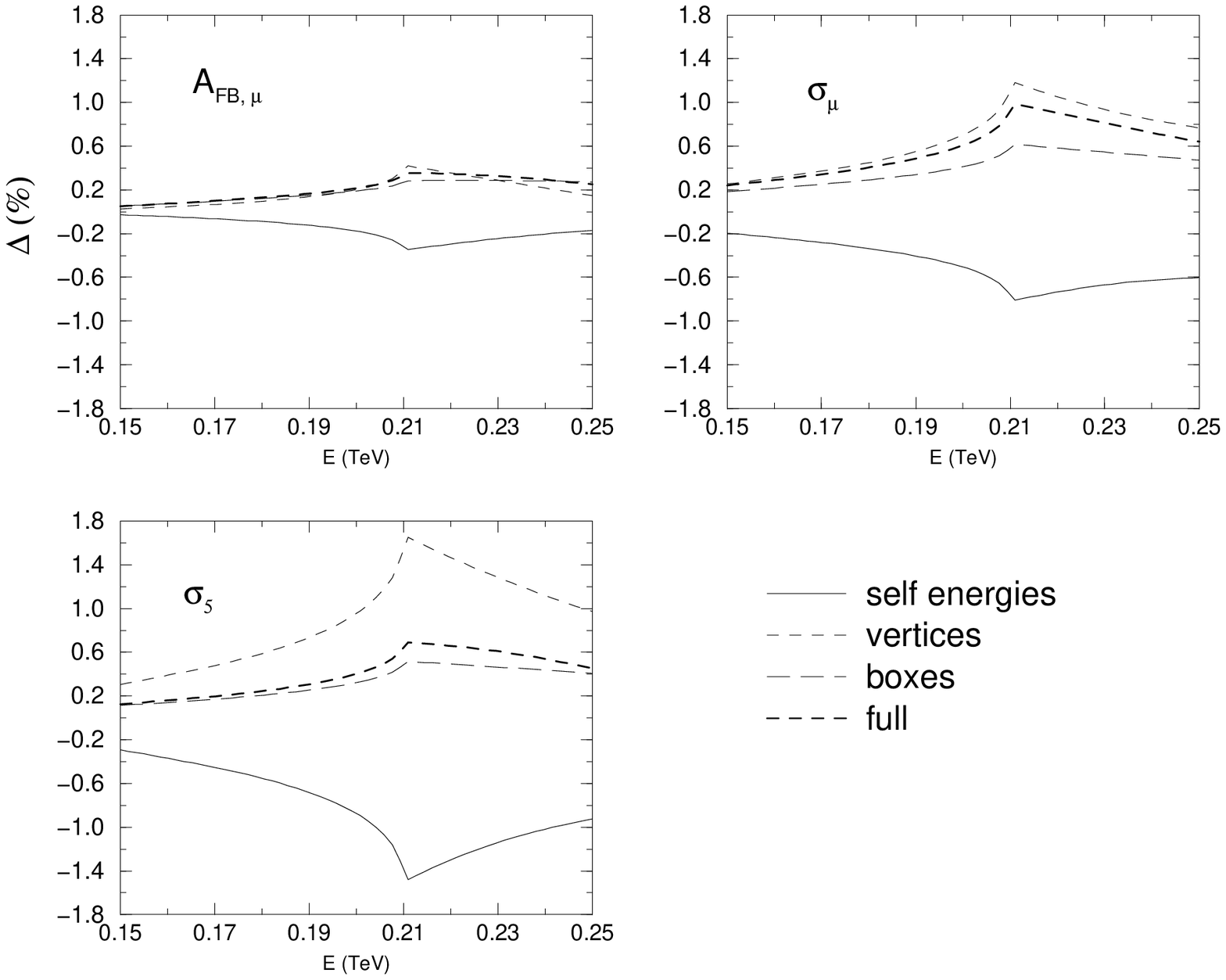, width=12cm}}
\caption{Selfenergy, box and vertex
SUSY effects in the heavy sfermions-light chargino scenario (same
parameters as in Fig.1 with a high $\mu$ value).}
\label{lightchar}
\end{figure}

\newpage
\begin{figure}[htb]
\vspace{15cm}
\includegraphics{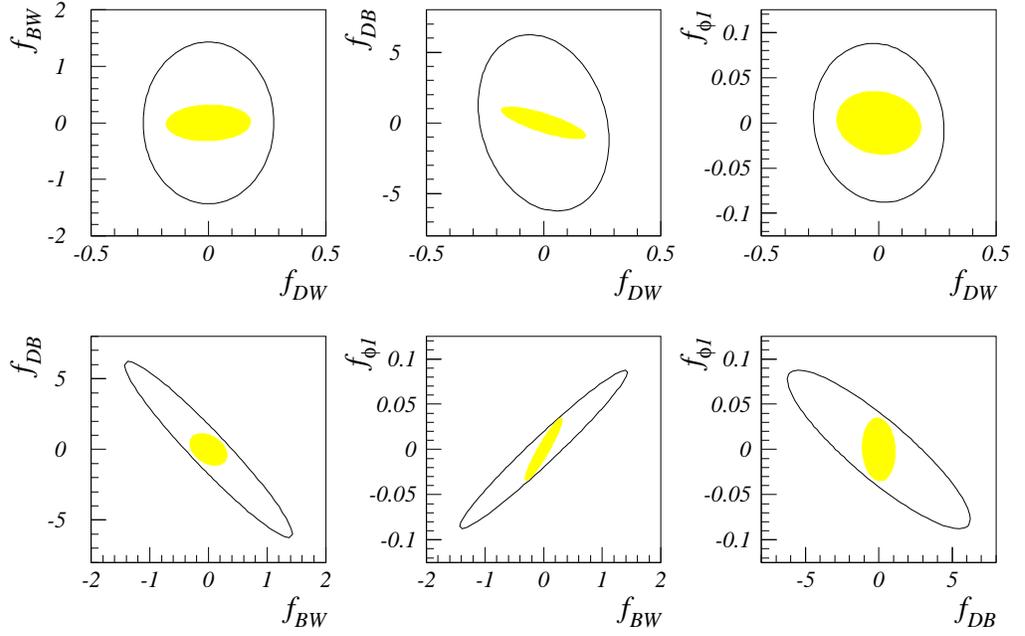}
\caption{Projection of the $\chi^2<\chi^2_{min}+1$ region in the 
space of the four anomalous gauge couplings onto the six possible 
coordinate planes. The outer ellipses are obtained from low energy data only;
the inner ones, by including the LEP2 data.}
\label{low-high}
\end{figure}

\clearpage
\newpage

\begin{figure}[p]
\vspace*{-4cm}
\[
\epsfig{file=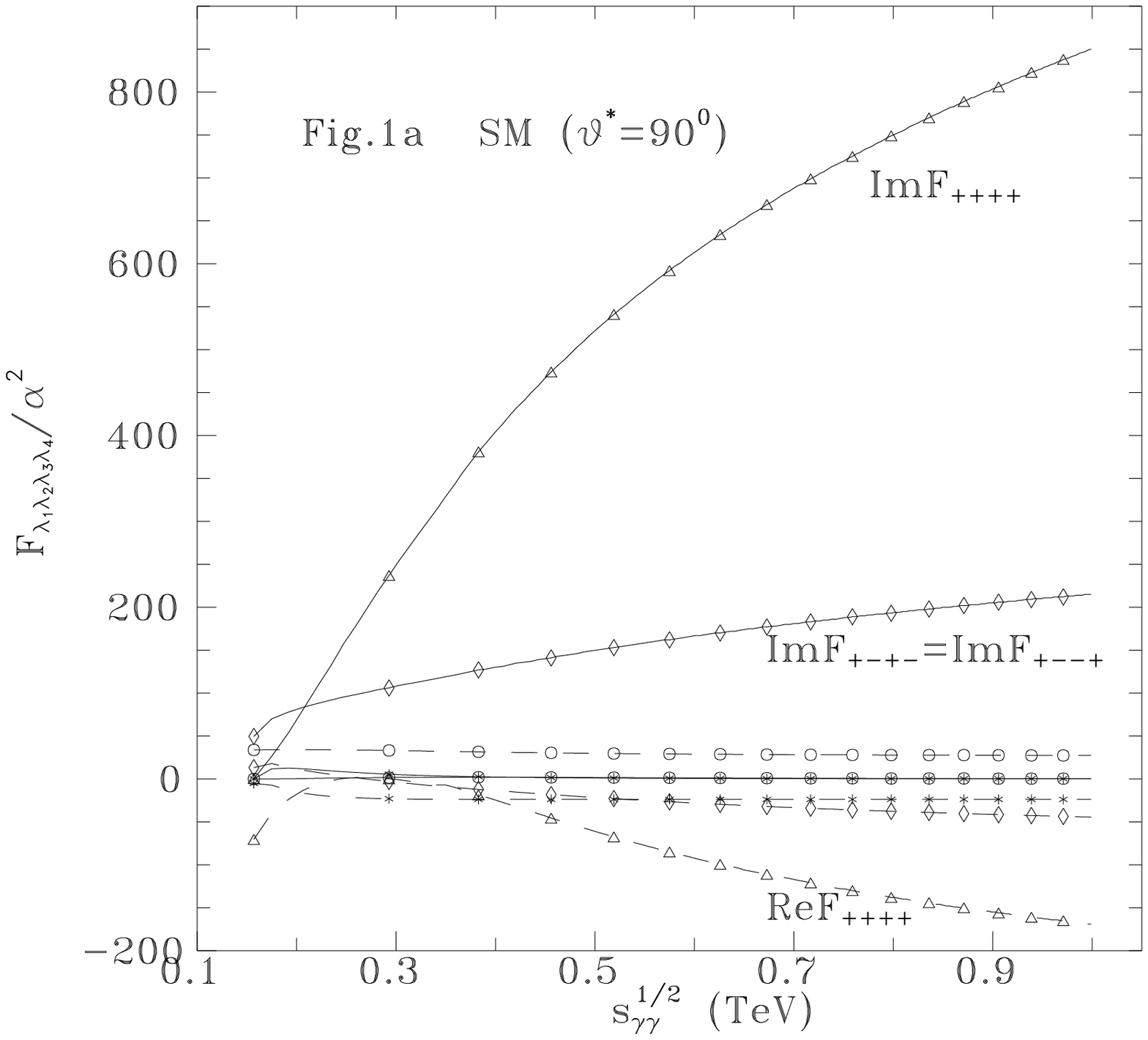,height=7.5cm}\hspace{0.5cm}
\epsfig{file=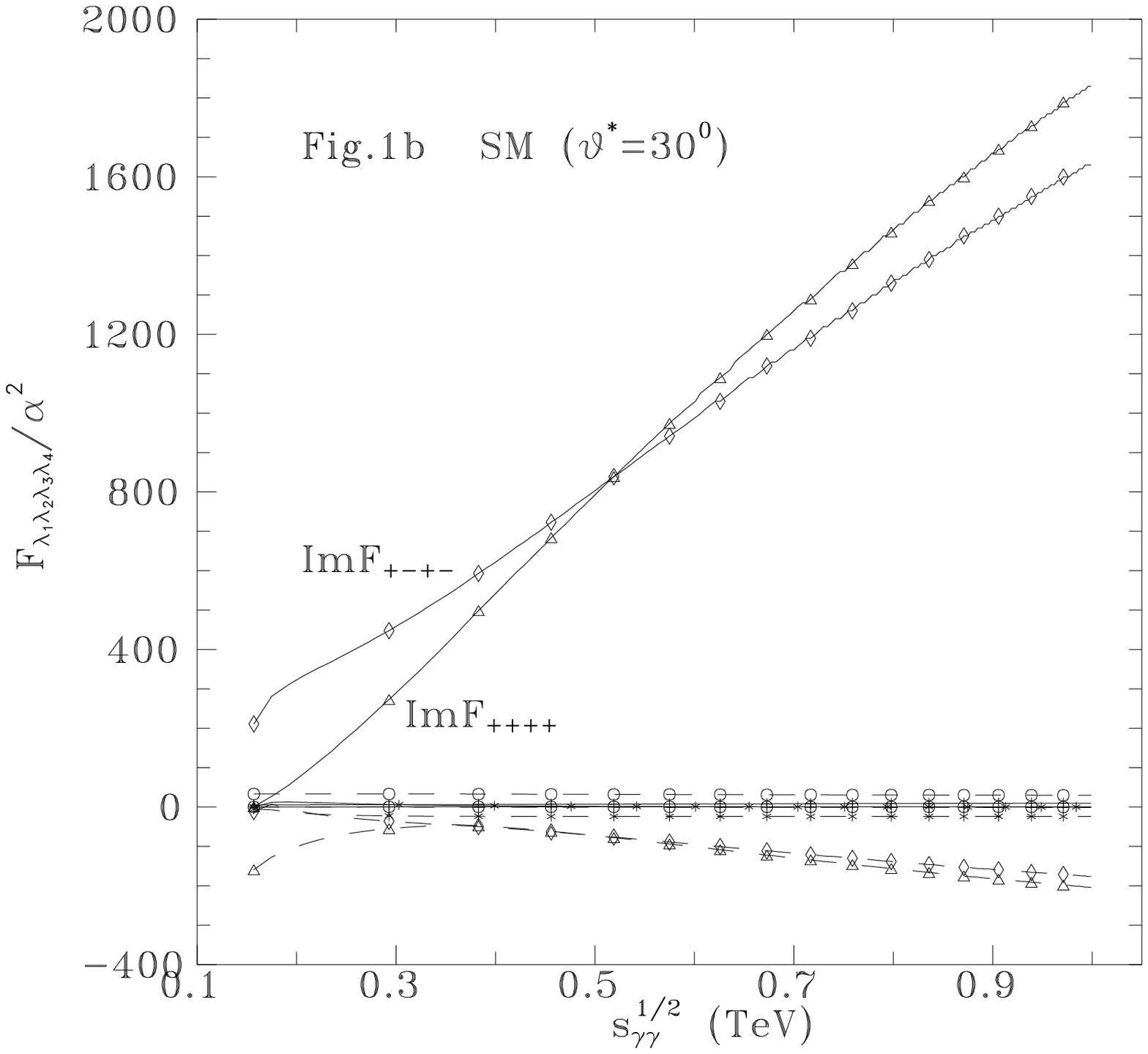,height=7.5cm}
\]
\hspace{8.3cm}$\gamma \gamma \to \gamma \gamma $
\vspace*{1cm}
\[
\epsfig{file=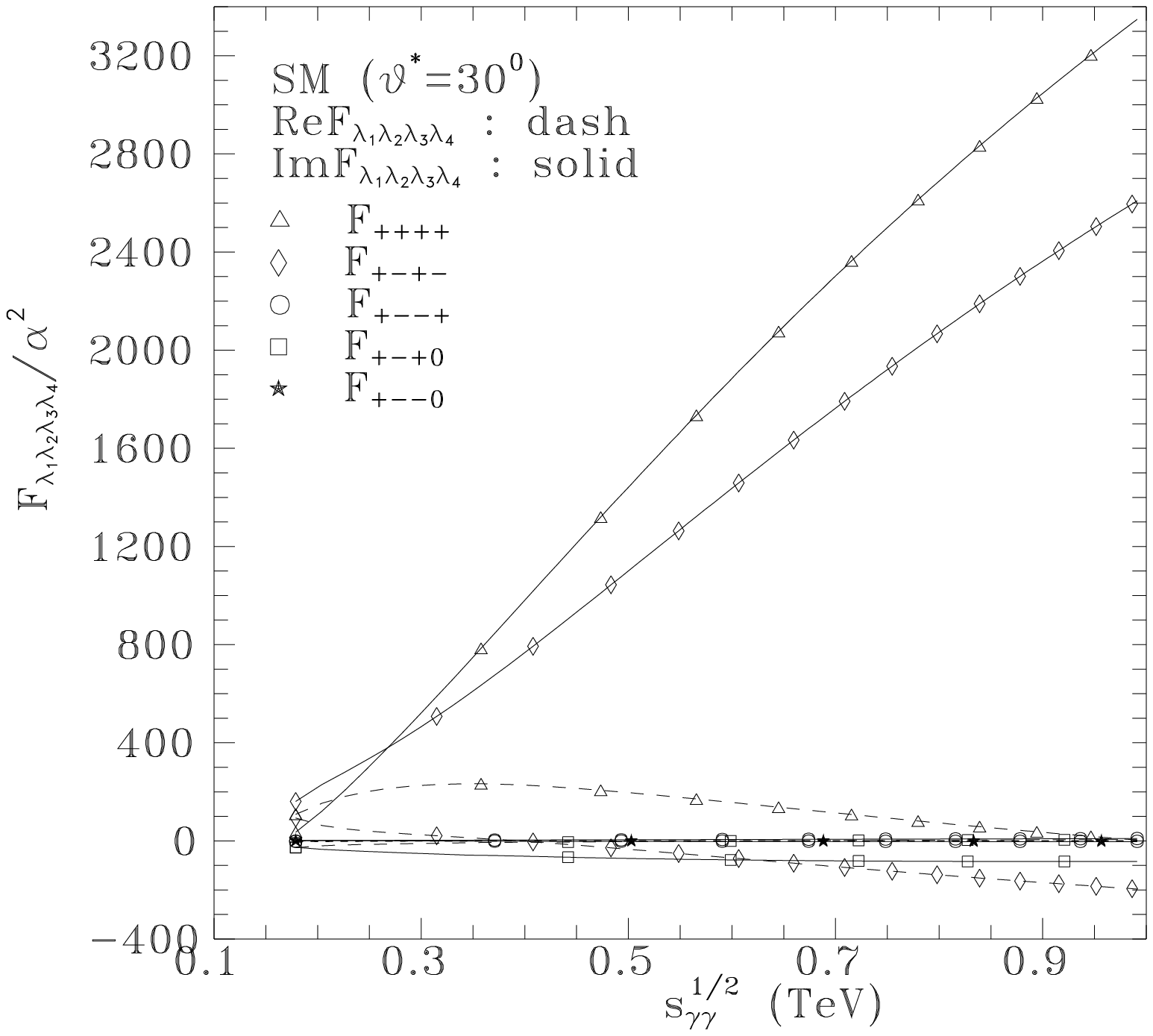,height=7.5cm}\hspace{0.5cm}
\epsfig{file=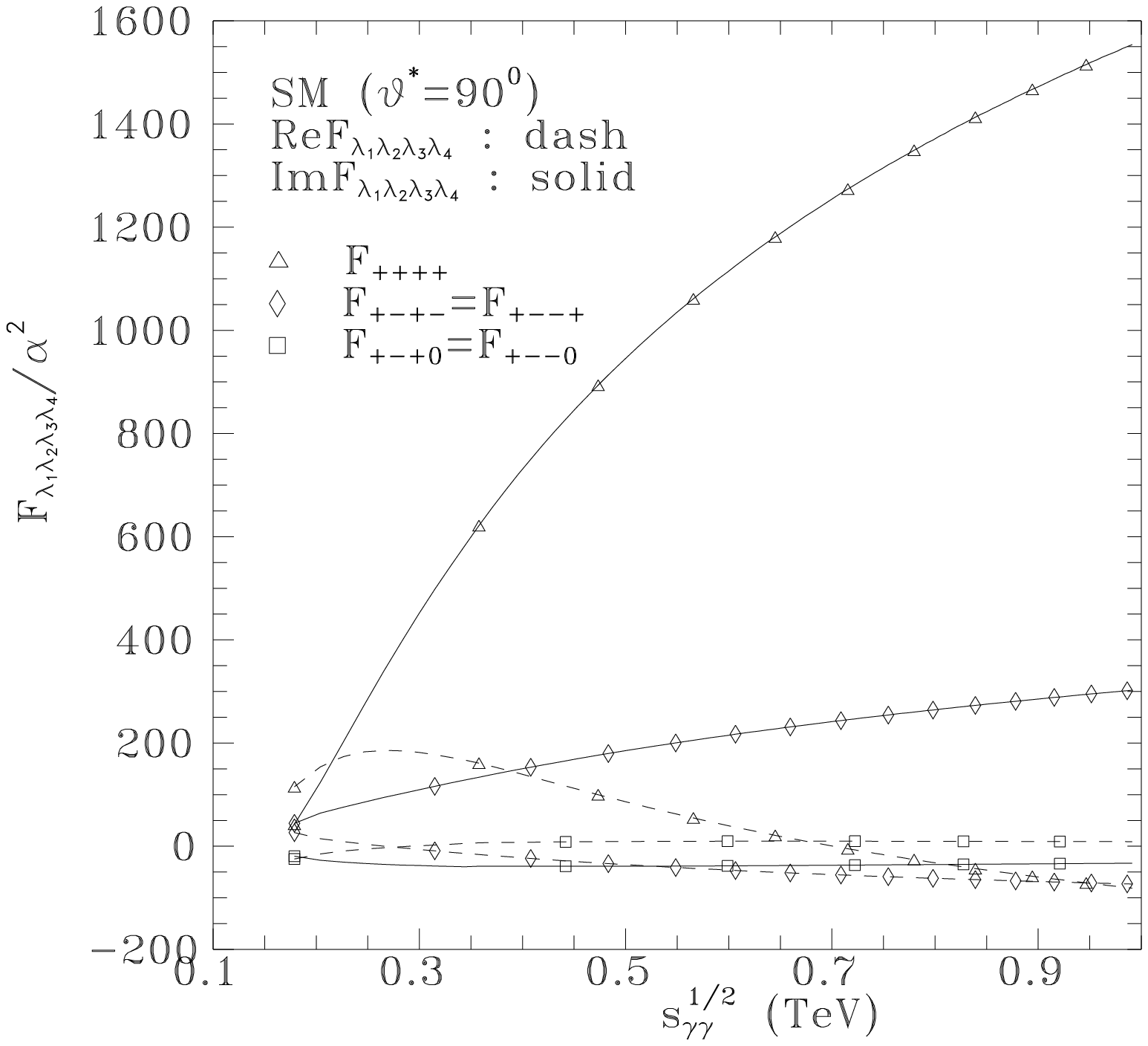,height=7.5cm}
\]
\hspace{8.5cm}$\gamma \gamma \to \gamma Z $
\caption[1]{Imaginary (solid line) and real (dashed line) parts
of the SM $\gamma \gamma \to \gamma \gamma $ 
and $\gamma \gamma \to \gamma Z $
helicity amplitudes
at $\vartheta =90^0$  and $\vartheta =30^0$.}
\label{ampSM}
\end{figure}

\clearpage
\newpage

\begin{figure}[p]
\vspace*{-4cm}
\[
\epsfig{file=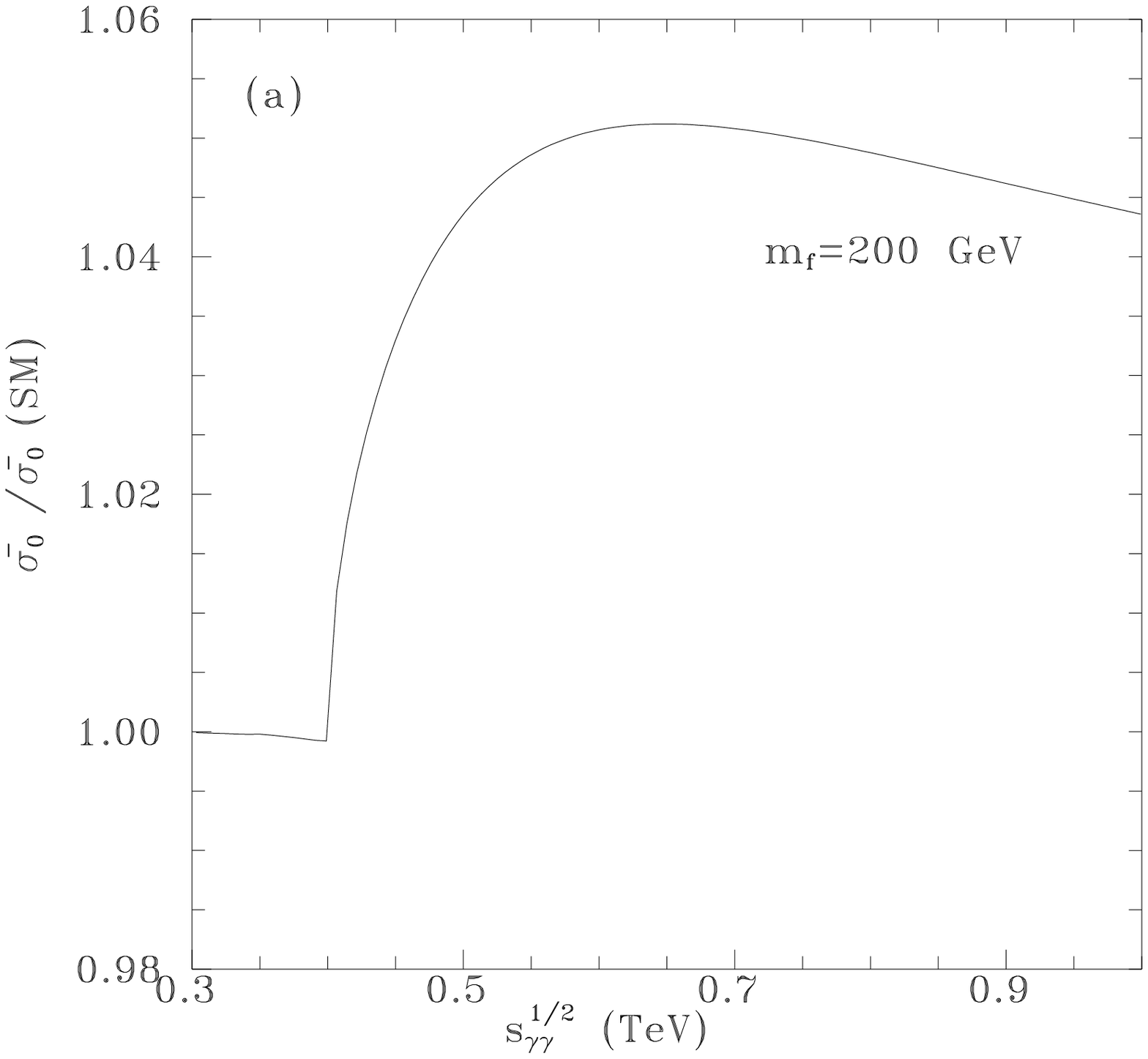,height=7.5cm}\hspace{0.5cm}
\epsfig{file=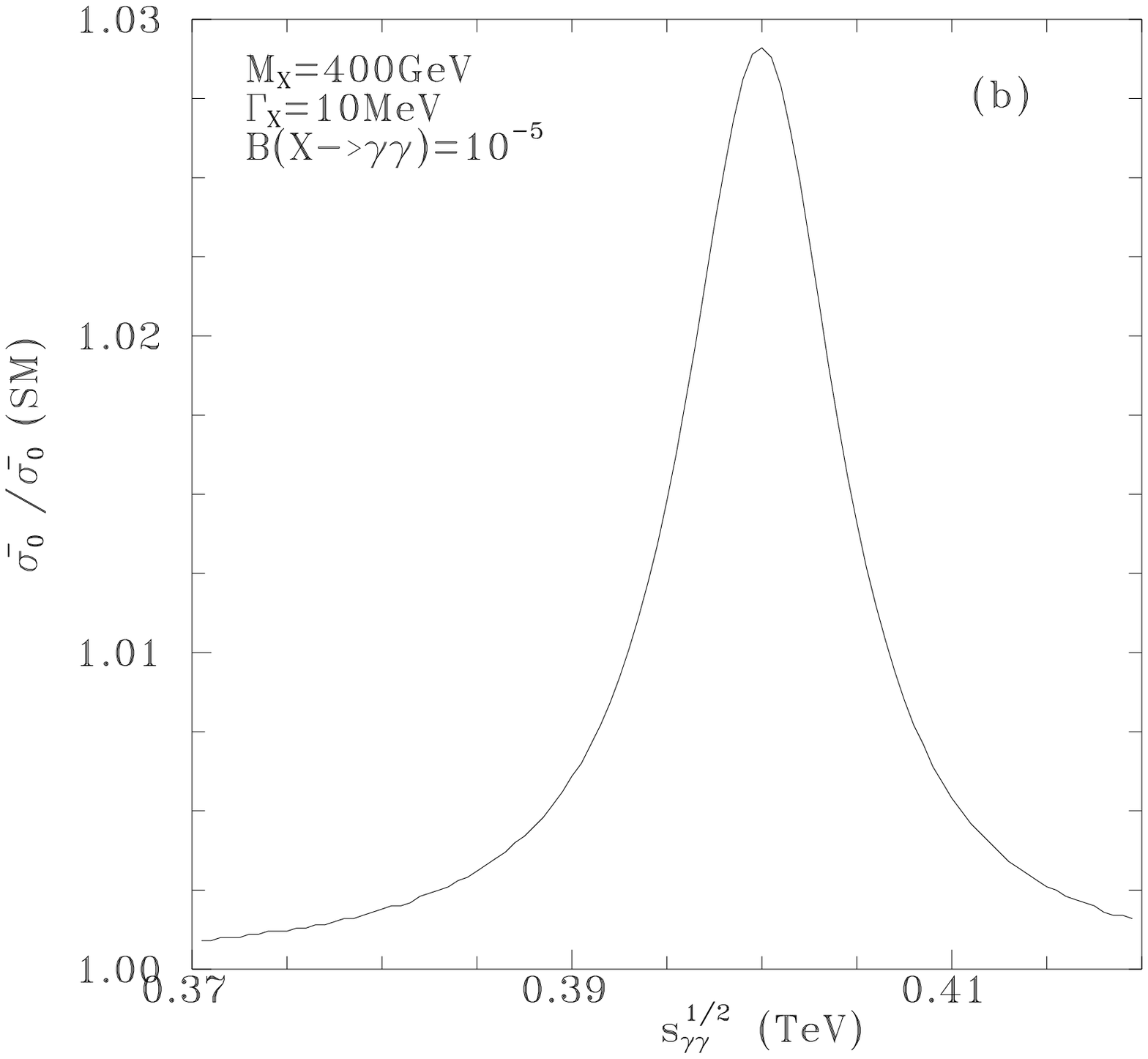,height=7.5cm}
\]
\vspace*{1.5cm}
\[
\epsfig{file=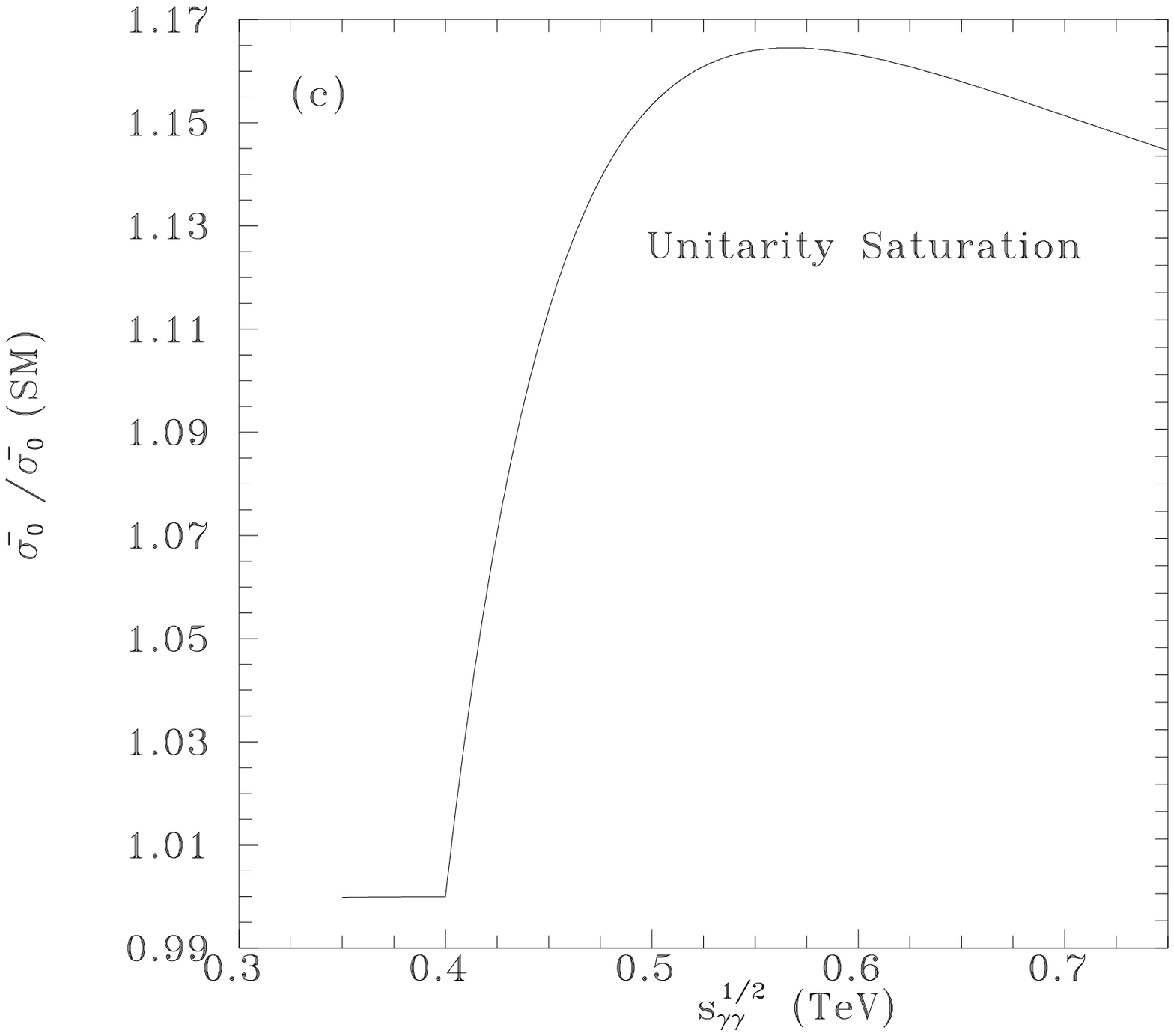,height=7.5cm}
\]
\vspace*{1.cm}
\caption[1]{Relative magnitude with respect to the SM results,
of the unpolarized $\gamma \gamma \to \gamma \gamma $
cross sections for a charge $+1$ fermion (a),
a typical s-channel neutral resonance (b), and 
unitarity saturating amplitudes (c). In all cases 
the cross sections have been integrated in the c.m. angular range
$30^0 < \vartheta^* < 150^0$. }
\label{models}
\end{figure}

\clearpage
\newpage

\begin{figure}[p]
\vspace*{-3cm}
\[
\epsfig{file=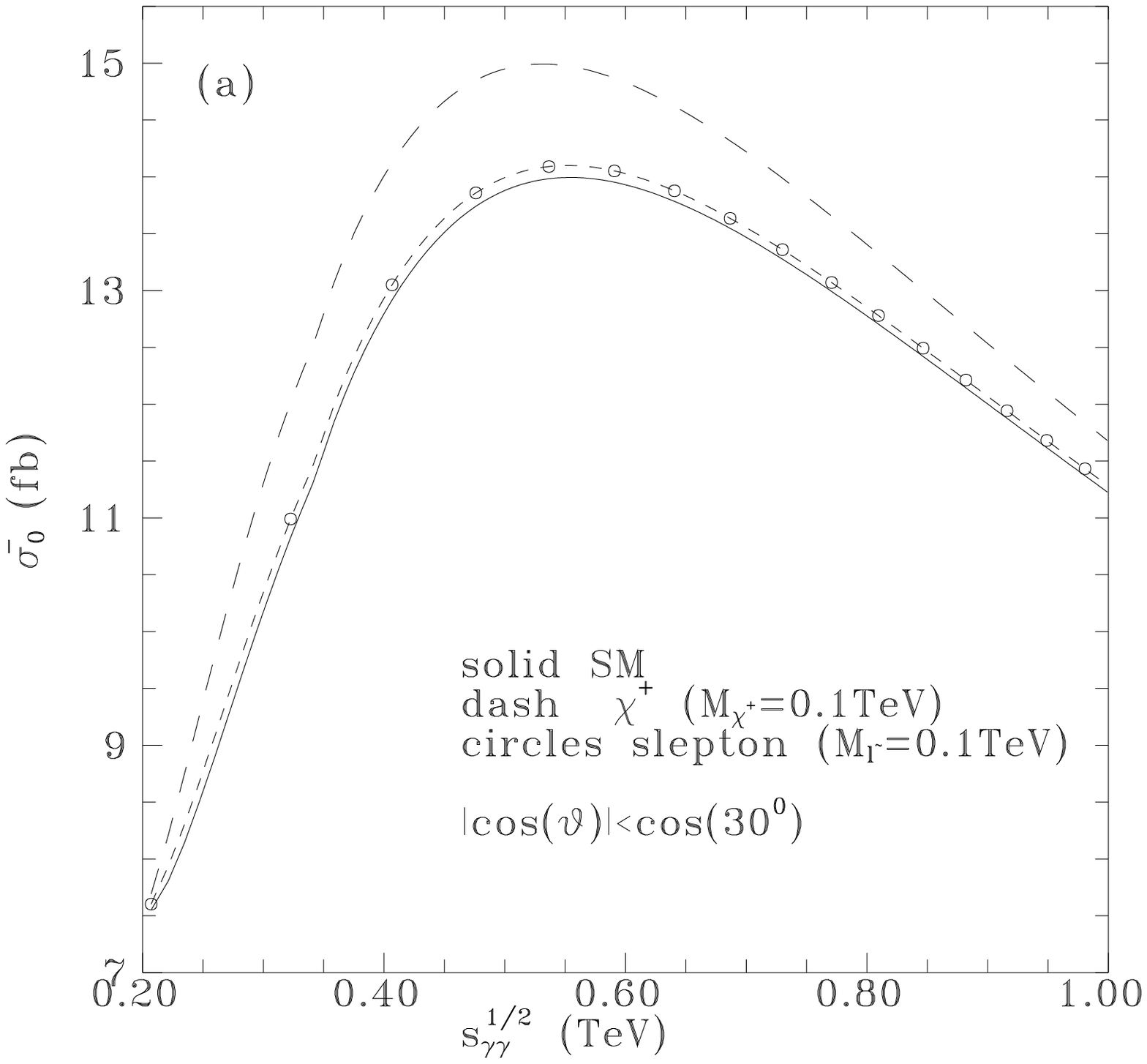,height=7.5cm}\hspace{0.5cm}
\epsfig{file=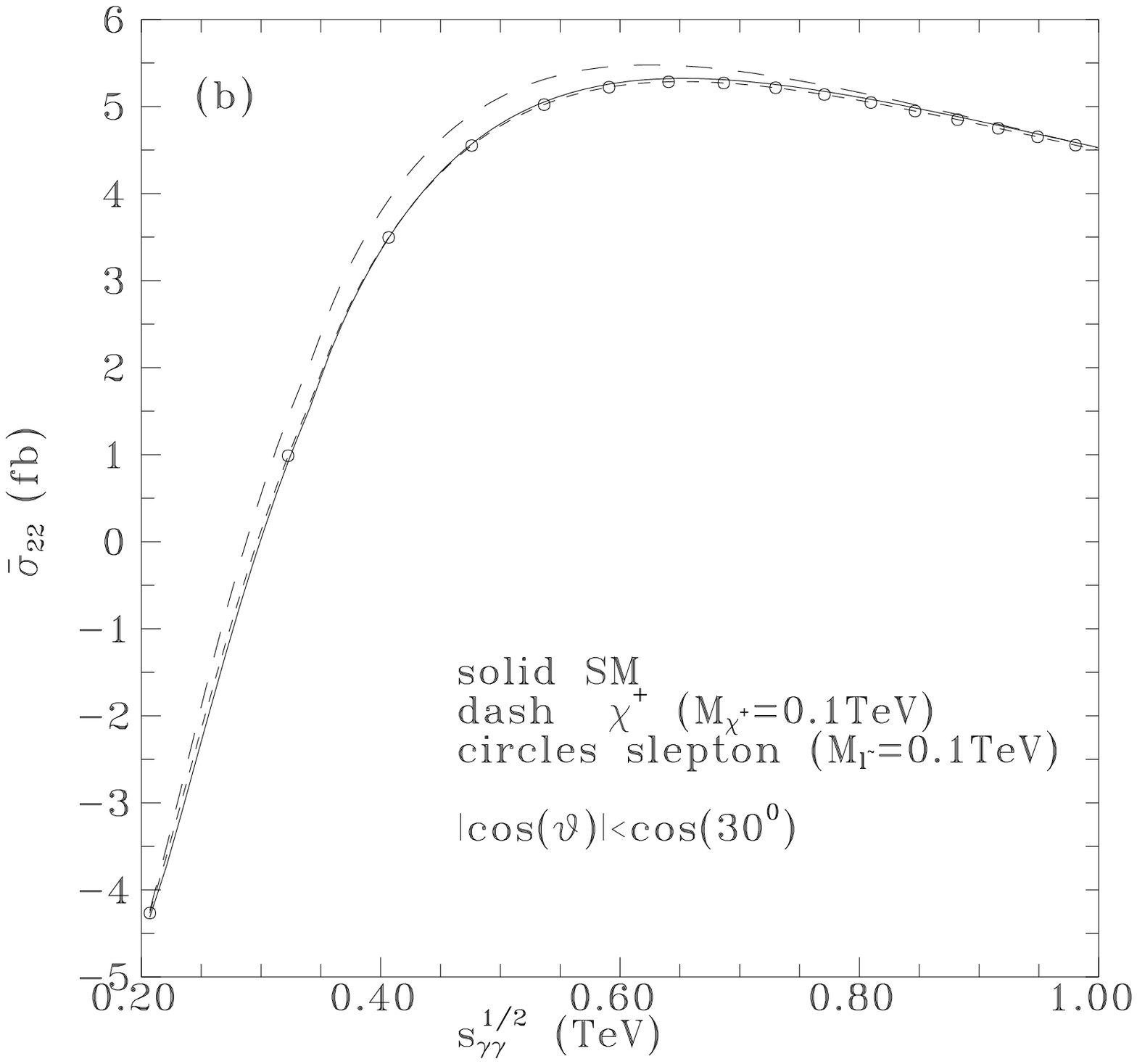,height=7.5cm}
\]
\vspace*{1.5cm}
\[
\epsfig{file=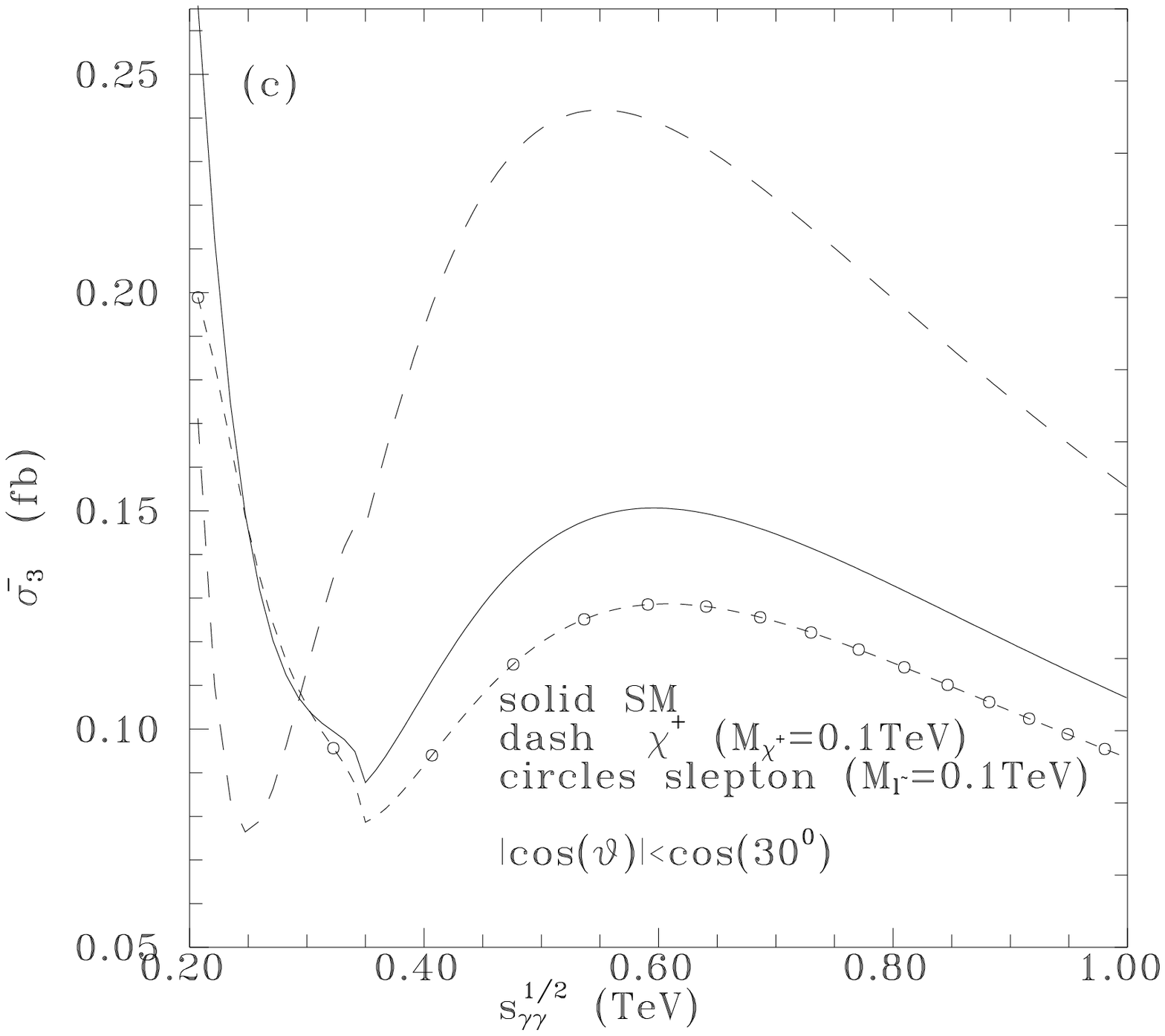,height=7.5cm}\hspace{0.5cm}
\epsfig{file=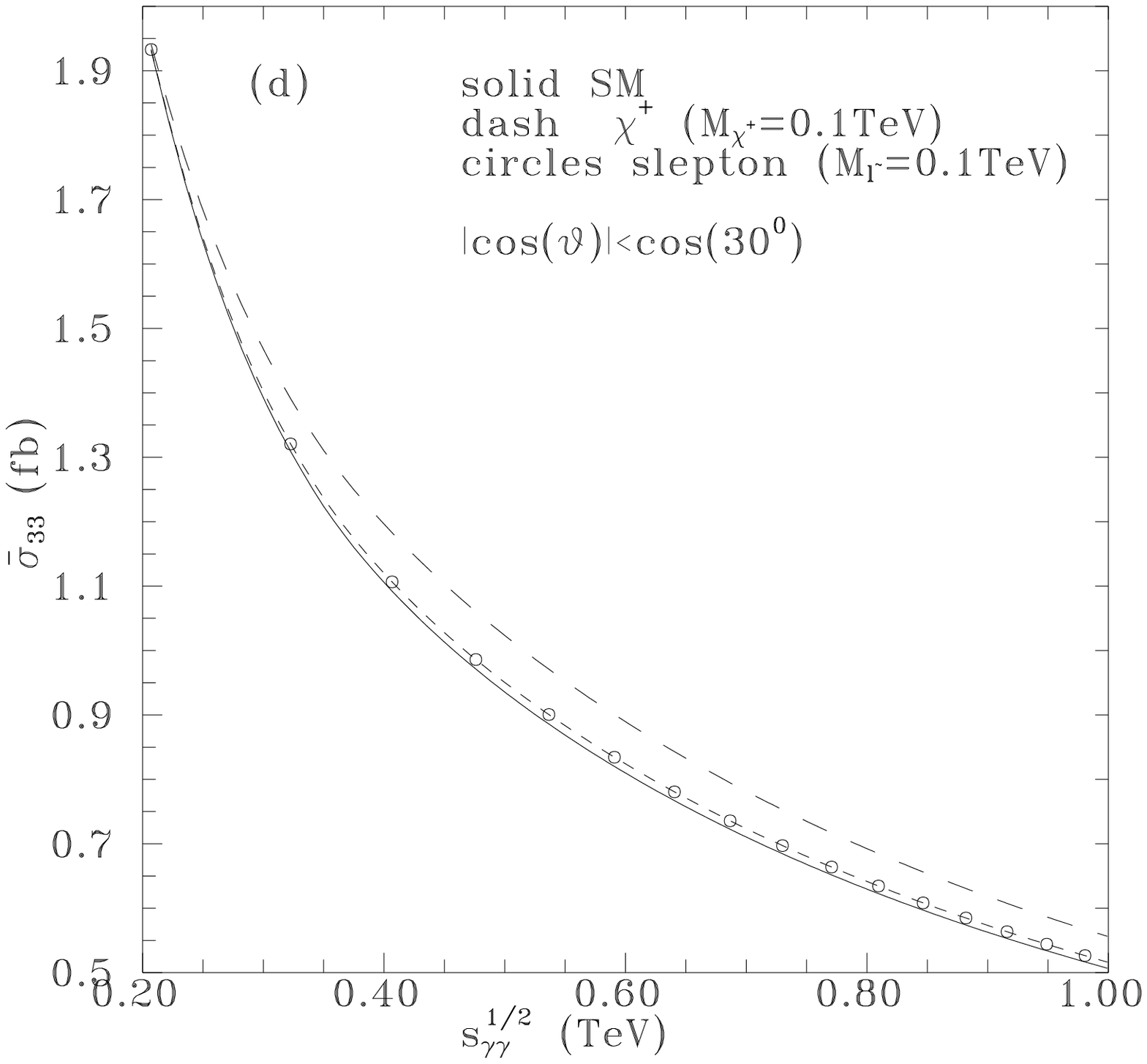,height=7.5cm}
\]
\vspace*{1.cm}
\caption[1]{$\bar \sigma_0$, $\bar \sigma_{22}$, $\bar \sigma_{3}$,
$\bar \sigma_{33}$ 
cross sections in $\gamma\gamma\to\gamma \gamma$
integrated over
$|\cos(\vartheta^*)|<\cos(30^0)$.
The SM and SUSY contributions induced by one chargino or
one charged slepton with mass
of 100 GeV, are also indicated.}
\label{SUSYgg}
\end{figure}

\clearpage
\newpage

\begin{figure}[p]
\vspace*{-3cm}
\[
\epsfig{file=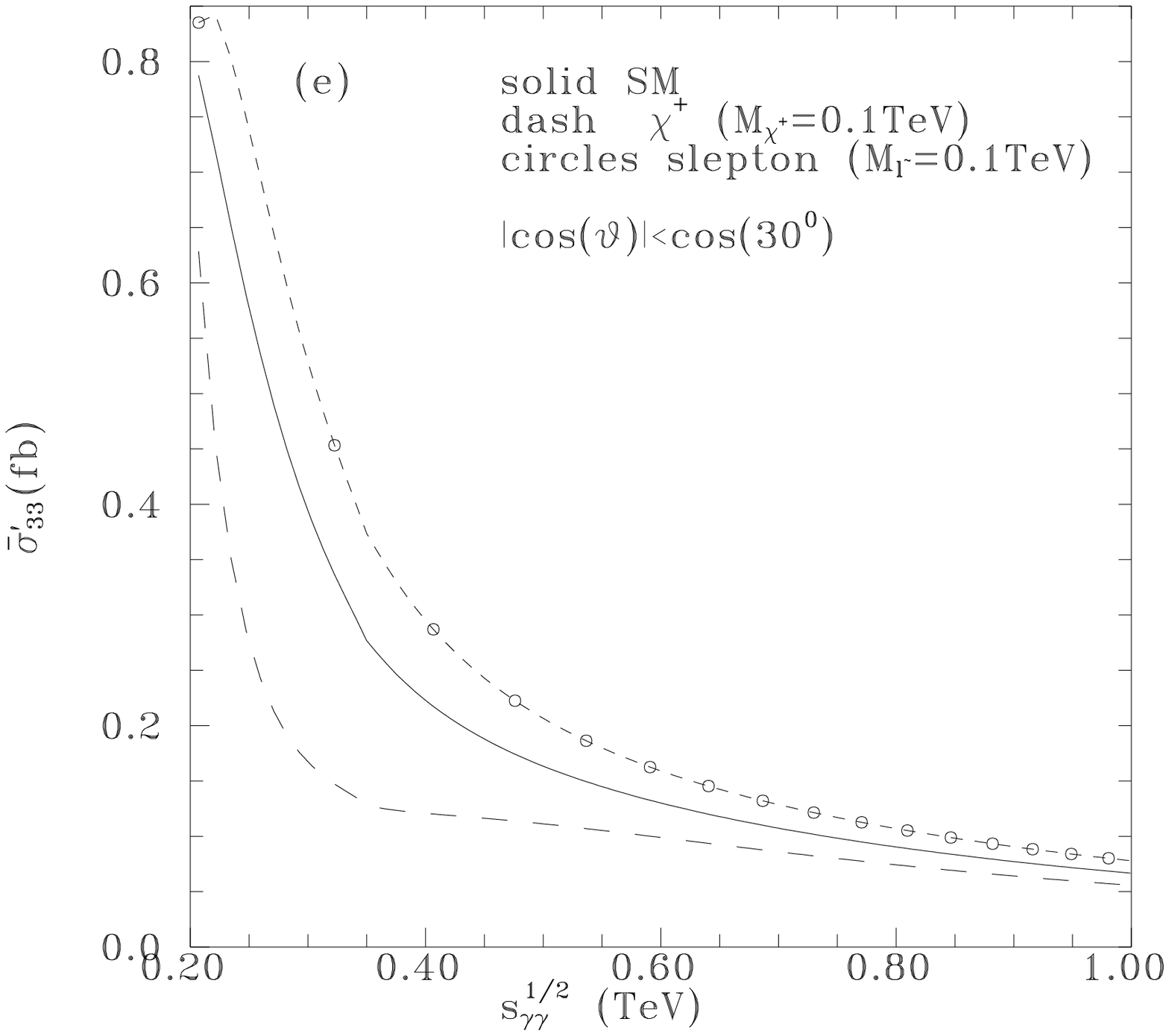,height=7.5cm}\hspace{0.5cm}
\epsfig{file=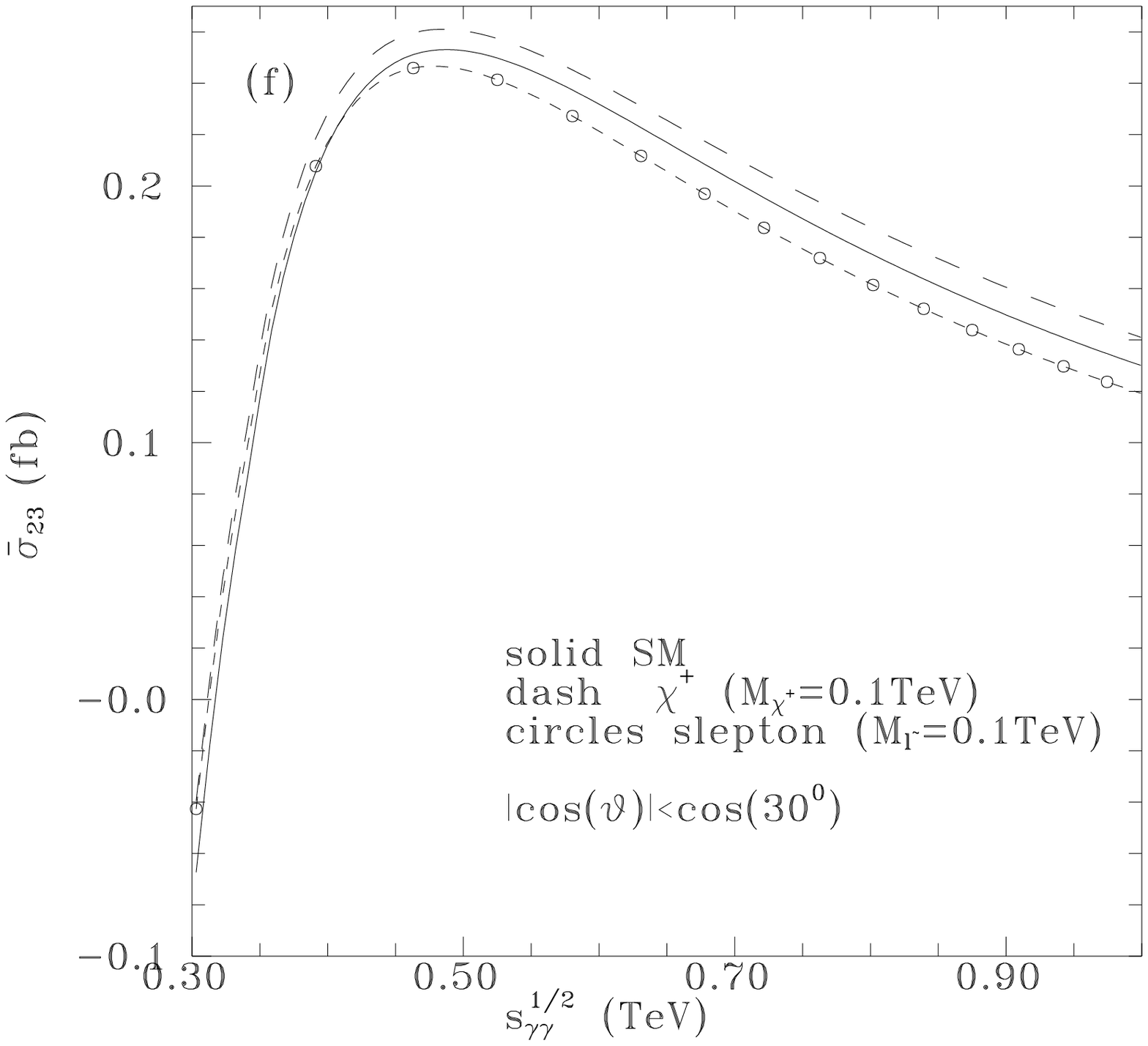,height=7.5cm}
\]
\vspace*{1.cm}
\caption[1]{$\bar \sigma'_{33}$  and
$\bar \sigma_{23}$ cross sections in $\gamma\gamma\to\gamma
\gamma$ integrated over
$|\cos(\vartheta^*)|<\cos(30^0)$
The SM and SUSY contributions induced by one chargino or
one charged slepton with mass
of 100 GeV, are also indicated.}
\label{SUSYgg1}
\end{figure}

\clearpage
\newpage

\begin{figure}[p]
\vspace*{-3cm}
\[
\epsfig{file=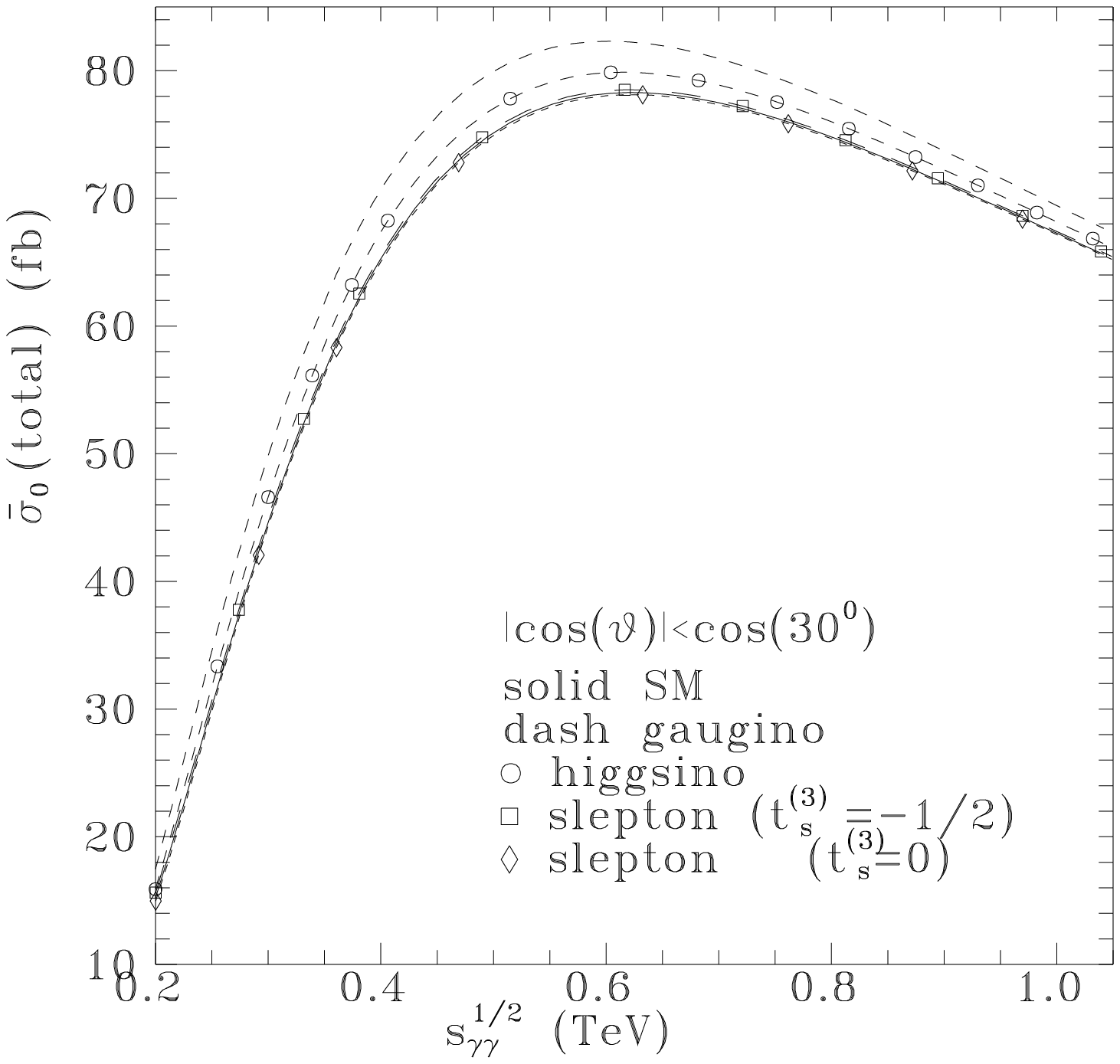,height=7.5cm}\hspace{0.5cm}
\epsfig{file=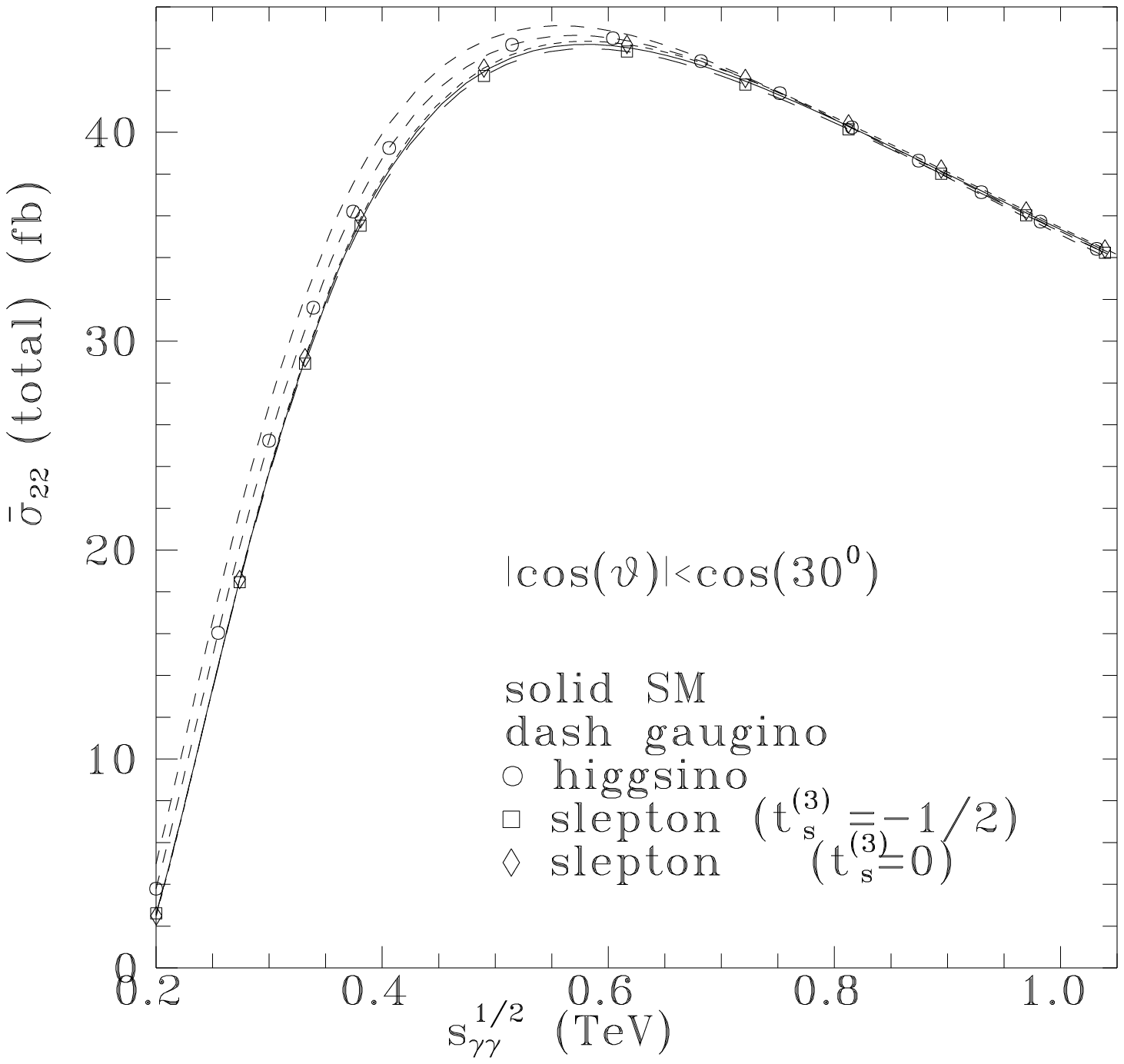,height=7.5cm}
\]
\vspace*{1.5cm}
\[
\epsfig{file=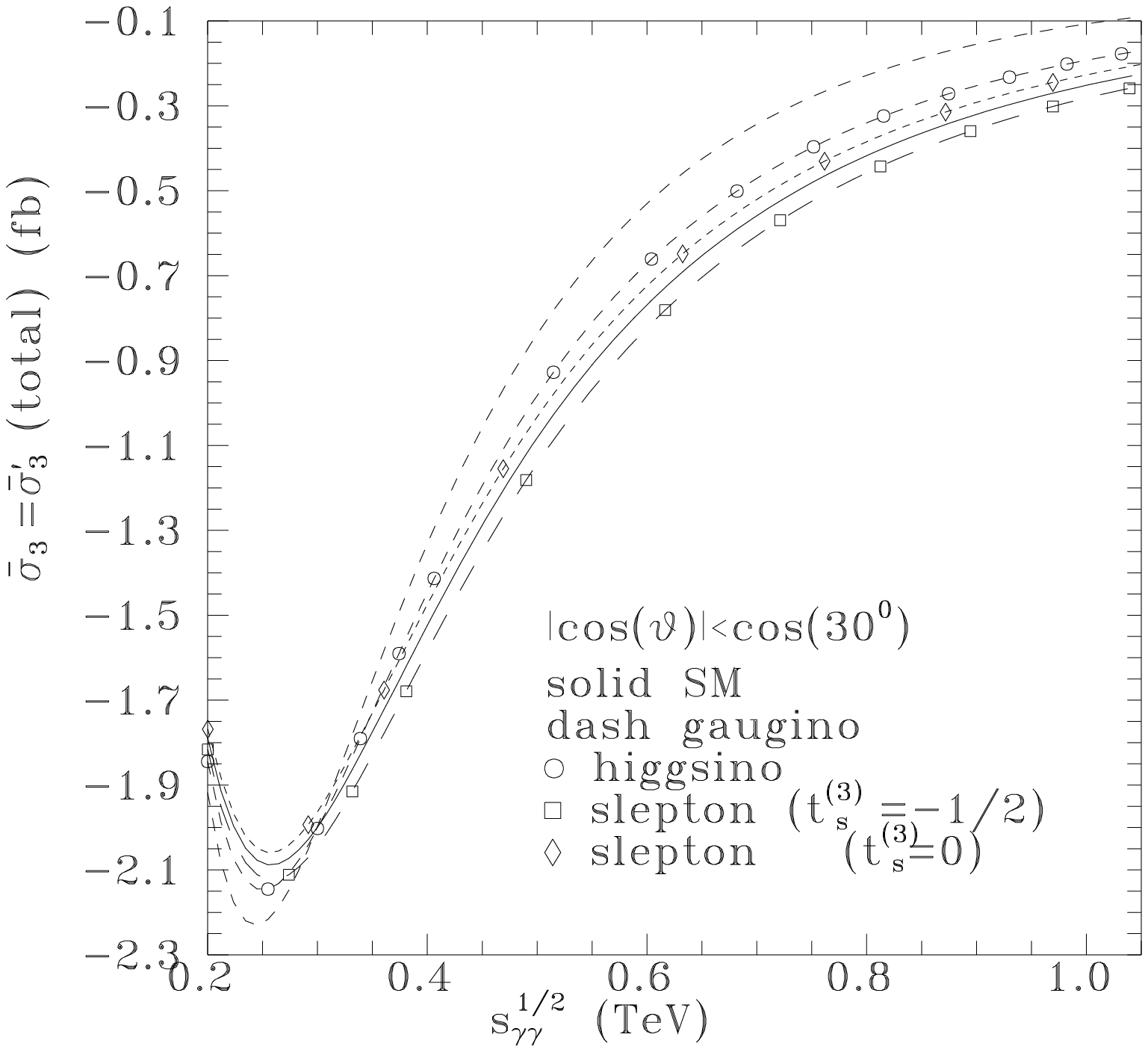,height=7.5cm}\hspace{0.5cm}
\epsfig{file=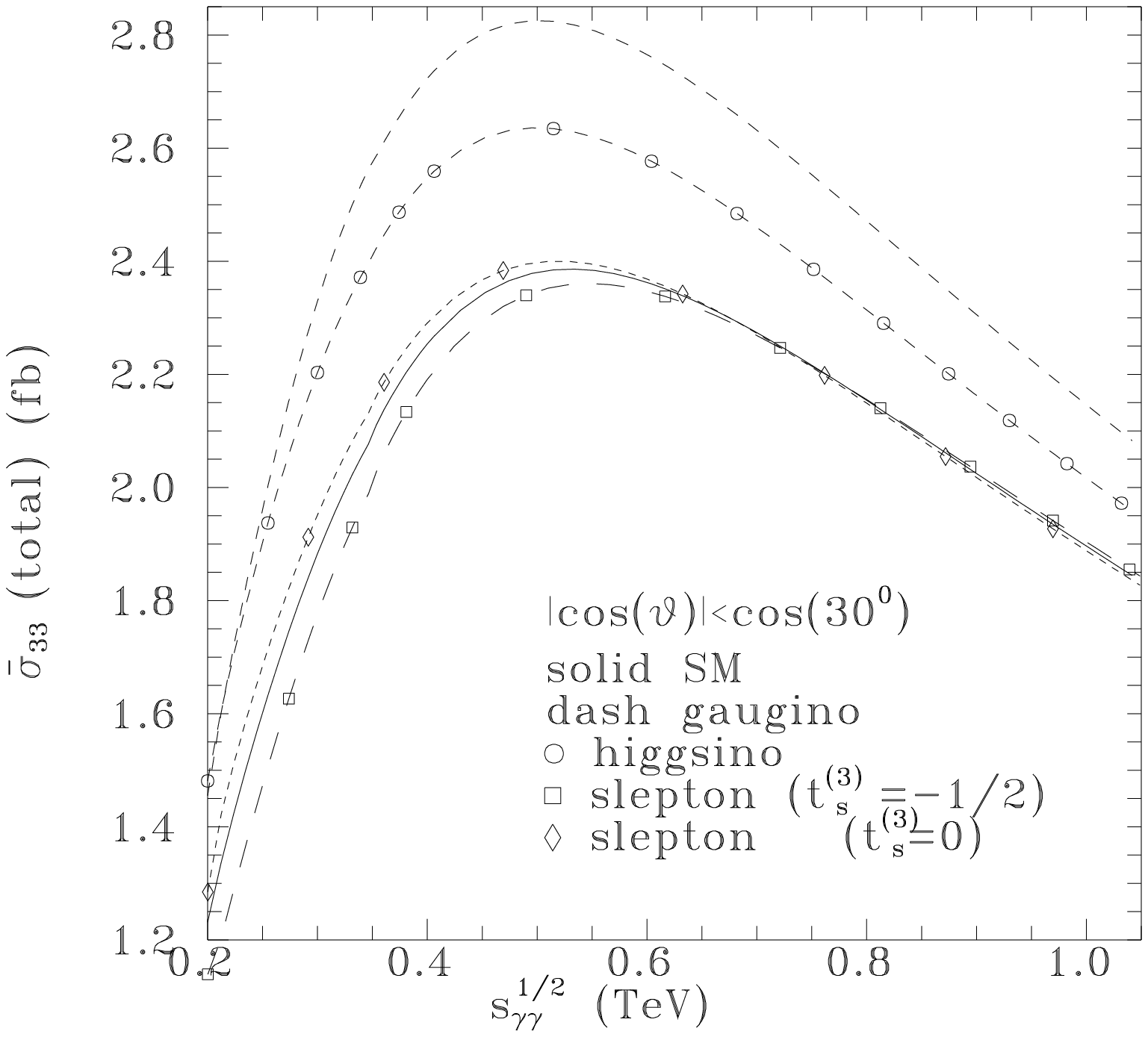,height=7.5cm}
\]
\vspace*{1.cm}
\caption[1]{$\bar \sigma_0$, $\bar \sigma_{22}$, $\bar \sigma_{3}$,
$\bar \sigma_{33}$ 
cross sections in $\gamma\gamma\to\gamma Z$ integrated over
$|\cos(\vartheta^*)|<\cos(30^0)$.
The SM and SUSY contributions induced by one chargino or
one charged slepton with mass
of 100 GeV, are also indicated.}
\label{SUSYgZ}
\end{figure}

\clearpage
\newpage

\begin{figure}[p]
\vspace*{-3cm}
\[
\epsfig{file=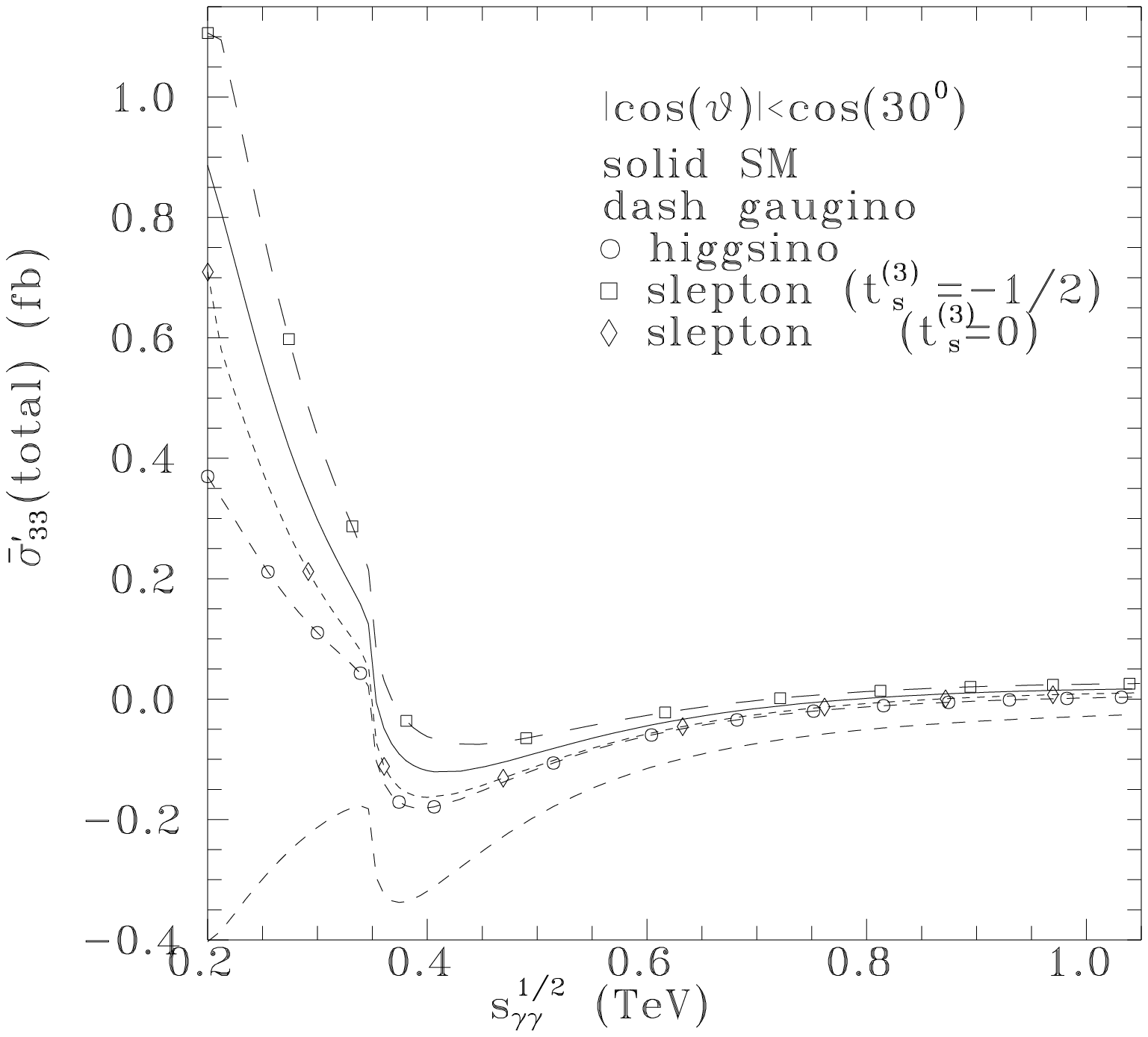,height=7.5cm}\hspace{0.5cm}
\epsfig{file=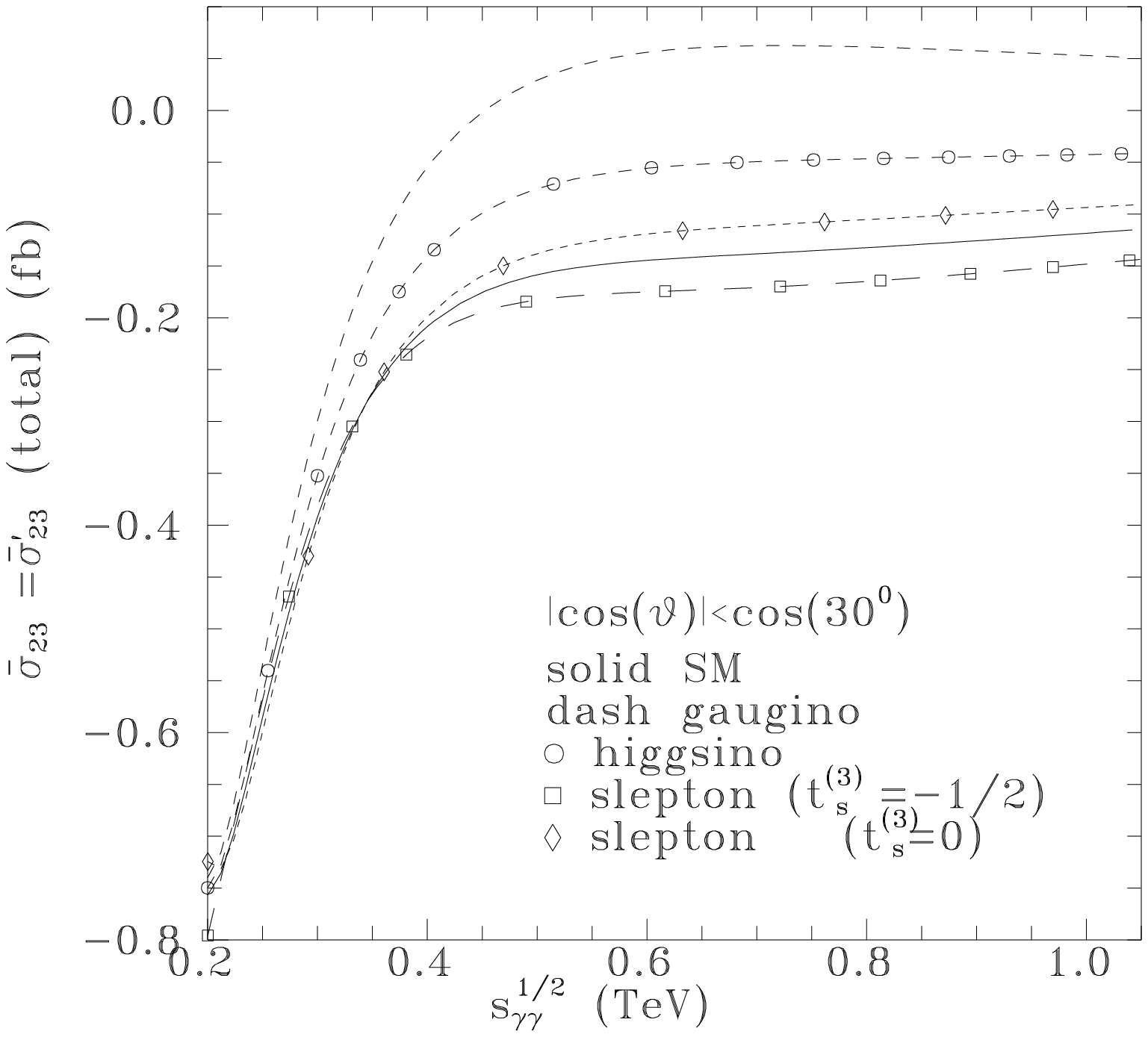,height=7.5cm}
\]
\vspace*{1.cm}
\caption[1]{$\bar \sigma'_{33}$  and
$\bar \sigma_{23}$ cross sections in $\gamma\gamma\to\gamma Z$
integrated over
$|\cos(\vartheta^*)|<\cos(30^0)$
The SM and SUSY contributions induced by one chargino or
one charged slepton with mass
of 100 GeV, are also indicated.}
\label{SUSYgZ1}
\end{figure}

\end{document}